\documentclass[aps,pra,twocolumn,showpacs,nofootinbib,longbibliography,notitlepage]{revtex4-2}
\usepackage{etex}
\usepackage{amsmath,amssymb,amsthm}
\usepackage[colorlinks=true,citecolor=blue,urlcolor=blue]{hyperref}
\usepackage[pdftex]{graphicx}
\usepackage{times,txfonts}
\usepackage{braket}
\usepackage{cleveref}
\usepackage{physics}
\usepackage{multirow}
\usepackage{verbatim}
\usepackage[T1]{fontenc}
\usepackage{color}
\usepackage[utf8]{inputenc}
\usepackage{accents}
\usepackage{xcolor}
\usepackage{natbib}
\usepackage{amsmath,blkarray}
\usepackage{mathtools}
\usepackage{latexsym}
\usepackage{tabularx, booktabs}
\usepackage{graphics,epstopdf}
\usepackage{float}
\usepackage{graphicx}
\usepackage{amsfonts}
\usepackage{caption}
\usepackage{rotating}
\usepackage{bigstrut}
\usepackage{color,soul}

\newcommand{\ba}{\begin{eqnarray}}
\newcommand{\ea}{\end{eqnarray}}

\newcommand{\X}{\sigma_{x}}

\newcommand{\Z}{\sigma_{z}}
\newcommand{\I}{\mathbb{I}}
\newcommand{\half}{\frac{1}{2}}
\newcommand{\sqt}{\frac{\sqrt{3}}{2}}
\newcommand{\qua}{\frac{1}{4}}

\begin{document}
	\title{Robust certification of unsharp instruments through sequential quantum advantages in a prepare-measure communication game}
	\author{Abhyoudai. S. S.}
 \affiliation{National Institute of Technology Patna, Ashok Rajpath, Patna, Bihar 800005, India}
	\author{Sumit Mukherjee}
	\email{mukherjeesumit93@gmail.com}
	\affiliation{National Institute of Technology Patna, Ashok Rajpath, Patna, Bihar 800005, India}
 \author{ A. K. Pan }
	\email{akp@phy.iith.ac.in}
	\affiliation{National Institute of Technology Patna, Ashok Rajpath, Patna, Bihar 800005, India}
 \affiliation{Department of Physics, Indian Institute of Technology Hyderabad, Telengana-502284, India }
	\begin{abstract}

Communication games are one of the widely used tools that are designed to demonstrate quantum supremacy over classical resources. In that, two or more parties collaborate to perform an information processing task to achieve the highest success probability of winning the game. We propose a specific two-party communication game in the prepare-measure scenario that relies on an encoding-decoding task of specific information. We first demonstrate that quantum theory outperforms the classical preparation non-contextual theory, and the optimal quantum success probability of such a communication game enables the semi-device-independent certification of qubit states and measurements. Further, we consider the sequential sharing of quantum preparation contextuality and show that, at most, two sequential observers can share the quantum advantage. The sub-optimal quantum advantages for two sequential observers form an optimal pair which certifies a unique value of the unsharpness parameter of the first observer. Since the practical implementation inevitably introduces noise, we devised a scheme to demonstrate the robust certification of the states and unsharp measurement instruments of both the sequential observers.
	\end{abstract}
	\pacs{03.65.Ta} 
	\maketitle
	\section{Introduction}
Arguably, Bell's theorem \cite{bell} has left the most prominent mark on the history of quantum foundations research. It states that all the quantum statistics cannot be predicted by a classical model that satisfies locality. This fundamental feature is known as quantum nonlocality, demonstrated through the quantum violation of Bell's inequality \cite{brunnerrev}. The nonlocal correlation that Bell's theorem certifies is device-independent,
i.e., the internal functioning of the quantum instruments remains unknown. This is the key reason why nonlocal correlations are used as a resource in various information-theoretic tasks, viz., secure quantum key distribution \cite{bar05,acin06,acin07,pir09},
randomness certification \cite{col06,pir10,nieto,col12}, witnessing Hilbert space dimension \cite{wehner,gallego,ahrens,brunnerprl13,bowler,sik16prl,cong17,pan2020}, and communication complexity \cite{complx1,buhrman16}.

Another aspect of non-classicality,  the inconsistency between the non-contextual hidden variable model and the quantum theory, was introduced by Kochen and Specker (KS) \cite{KSpeker68,bell66}. Unlike Bell's theorem, which requires spatially separated entangled systems to demonstrate nonlocality, the KS theorem can reveal the contextuality even using a single system. However, the KS theorem has a limited scope of applicability as it only captures the measurement non-contextuality, is not applicable for unsharp measurement, and is restricted to quantum theory. Later, the notion of non-contextuality was generalized in \cite{spekkens05} for arbitrary operational theories and extended the formulation to preparation and transformation. Of late, the preparation contextuality has been studied extensively\cite{spek09,hameedi,ghorai,saha19b,pan19,kumari2019,mukherjee22,flatt22,schmid18,lostaglio20}, and it is this form of non-classicality that plays a vital role in the present work. In particular, we show how quantum preparation contextuality powers a sequential communication game and enables the robust certification of states, observables, and unsharpness parameters.
	
The strongest form of certification is device-independent self-testing \cite{supicrev,kaniewski}, whereby one can uniquely certify the states and measurements solely from experimental statistics regardless of any knowledge about the functioning of the preparation and measurement instruments. Optimal quantum violation of a Bell inequality enables such self-testing. For instance, the optimal quantum violation of Clauser-Horne-Shimony-Holt \cite{chsh} inequality self-tests the maximally entangled state and locally anti-commuting observables. However,  in practice, the loophole-free Bell tests remain a challenging and resource-consuming task. Instinctively, robust certification of states and measurement instruments through the Bell test would be even more challenging. Due to the practical difficulties in experimental testing, the semi-device-independent (SDI) protocols  in prepare-measure scenario\cite{pawlowski11,tava2018,farkas2019,tava21,miklin2020,anwar2020,fole2020,tava20exp,sumit21} became appealing to the community. In the SDI certification scheme, the preparation and measurement instruments remain black boxes, but the dimension of the system is bounded from above. 
	
One of the well-studied approaches for showcasing quantum supremacy in information processing tasks is in terms of communication games \cite{buhrman10,buhrman16, spek09,hameedi,tava2020}. A typical communication game involves two or more independent parties collaborating to perform an information processing task with the maximum probability of success. Apart from structural differences, such games are supplemented with different constraints that have to be fulfilled by the parties. In this work, we consider a particular communication game constrained by parity-oblivious communication, which implies that the communication between the parties must not reveal the parity-information of the input. The game is technically inspired from but not equivalent to the parity-oblivious communication game \cite{spek09}. Here we demonstrate that quantum theory predicts the success probability that outperforms the classical theory. Throughout this paper, by classicality, we refer to the preparation non-contextuality, as will be introduced shortly. 

Recently, the sequential sharing of quantum correlations has attracted considerable attention. Based on the sequential quantum violation of CHSH inequality,  Silva \emph{et al.},\cite{silva2015} first demonstrated that the nonlocality can be shared by a maximum of two sequential observers on one side. Since then, numerous studies \cite{silva2015,sasmal2018,bera2018,kumari2019,brown2020,Zhang21,sumit21,mohan2019,tava21} have been reported that explored how many independent observers can share various forms of quantum correlations, such as steering \cite{sasmal2018}, entanglement \cite{bera2018}, non-locality \cite{brown2020}, and preparation contextuality \cite{kumari2019}. As long as a communication game is capable of revealing some aspect of quantum supremacy, there is a provision to extend the game to the sequential scenario. We adopt a sequential scenario of the communication game to present the main results, which are explicitly discussed below.
	
Based on a specific parity-oblivious communication game in the prepare-measure scenario, we first demonstrate the SDI certifications of the prepared qubit states, measurement, and the unsharp parameter in an ideal scenario when there is no noise. The sequential quantum advantages over the classical preparation non-contextual bound of the communication game enable us to demonstrate such a certification. We show that, at most, two independent sequential observers can share the quantum advantage in such a game. Specifically, we argue how sub-optimal quantum advantages by two sequential observers form an optimal pair which eventually certifies a unique value of the unsharpness parameter of the first observer. 

However, the practical implementation of the protocol introduces inevitable losses due to imperfections in experimental procedures. Hence aforementioned optimal pair of sequential success probabilities may not be achieved. Thus, the exact states and measurements cannot be certified in an actual experiment. On the other hand, in a sequential scenario where more than one observer gets the quantum advantage, the success probabilities must be sub-optimal to ensure that the measurements are unsharp. We provide a robust certification of the preparations and measurements and certify a range of the unsharpness parameter of the measurement instrument when the sequential quantum success probabilities do not form the optimal pair. 

The paper is organized as follows. In Sec.\ref{sec:model} we briefly summarize the notion of the ontological model and the assumption of preparation noncontextuality.  In Sec.\ref{sec:POCG} we introduce the parity-oblivious communication game in the prepare-measure scenario. In Sec.\ref{sec:SQA}, using sequential unsharp measurements, we demonstrate the sharing of quantum advantage by multiple independent observers. In Sec.\ref{sec:CoL} we provide the certification of the unsharpness parameter, which is necessary to violate the preparation non-contextual bound of success probability, and in Sec.\ref{sec:IST} we provide a generalized certification statement for the preparations and unsharp measurements in an ideal scenario. Also, in Sec.\ref{sec:Debbie} we examined the possibility of sharing the sequential quantum advantage with a third observer. Further, in Sec.\ref{sec:RST} we present the robust certification scheme which robustly certifies Alice's preparations and Bob and Charlie's measurements to noise for the choice of unsharpness parameter. Finally, we summarize our results and conclude in Sec\ref{sec:SandD}.

\section{Ontological model and preparation noncontextuality}\label{sec:model}

To begin with, we briefly recapitulate the ontological model \cite{harrigen,ballentine,banik14} of an operational theory and the assumption of the preparation noncontextuality \cite{spekkens05} - the notion of classicality that we are considering here. An operational theory essentially involves a set of preparation procedures $\{P\}$ and a set of measurement procedures $\{M\}$, which is designed to predict the operational statistics by means of the probability $p(k|P, M)$ of obtaining a particular outcome $k$ given the experimental arrangements. In operational quantum theory, the preparation is represented by the density matrix $\rho$, and the measurement is, in general, represented by positive-operator-valued-measures (POVMs) $\{E_{k}\}$ satisfying $\sum_{k} E_{k}=\mathbb{I}$. The quantum probability $p(k|P, M)s$  is provided by the Born rule as $p(k|P, M)=Tr[\rho E_{k}]$.

	In general, the ontological model \cite{harrigen, ballentine,banik14} of an operational theory provides an objective description of the physical events compatible with the experimental statistics. A preparation procedure $P$ in an ontological model corresponds a probability distribution $\mu(\lambda|P)$ that satisfies, $\int_{\Lambda}\mu(\lambda|P) d\lambda =1$, for $\forall \lambda \in \Lambda$. The variable $\lambda \in \Lambda$ represents the ontic state of the system. Given a POVM element $E_{k}$, the ontic state variable $\lambda$ assigns a response function $\xi({k}|\lambda,E_{k})$ with  $\sum_{k}\xi(k|\lambda,E_{k})=1$, $\forall\lambda$.  Any ontological model that is consistent with quantum theory must reproduce the statistical prediction of the theory given by the Born rule as,
	\begin{equation}
		\int_{\Lambda}\mu(\lambda|\rho,P)  \xi(k|\lambda,M) d\lambda = Tr(\rho E_{k}). \label{eq:born}
	\end{equation}
For our purpose here, to introduce the notion of preparation noncontextuality in an ontological model we consider the operationally equivalent experimental procedures. Two different preparation procedures $P_{0}$ and $P_{1}$ are called operationally equivalent if no measurement can distinguish them. Mathematically it means that for any two operationally equivalent procedures $P_{0}$ and $P_{1}$ we must have, $p(k|P_{0}, M)=p(k|P_{1},M) $, $\forall M,k$. In quantum theory, a density matrix $\rho$ prepared through two distinct procedures $P_{0}$ and $P_{1}$ cannot be distinguished by any measurement and thus constitutes such operationally equivalent preparations.  As argued in \cite{spekkens05} that if two operationally equivalent preparations $P_{0}$ and $P_{1}$ for an operational theory are equivalently represented in the corresponding ontological model by the same ontic state probability distributions, i.e.,  
  \begin{equation}
 \forall M, \ \ p(k|P_{0}, M)=p(k|P_{1},M)	\Rightarrow \mu_{P_{0}}(\lambda|\rho)=\mu_{P_{1}}(\lambda|\rho),
 \end{equation} 
 then the model is called preparation noncontextual. This particular form of noncontextuality is taken as the notion of classicality in this paper. To avoid clumsiness, we drop $M$ from the response function for the rest of this paper, as measurement contextuality is not relevant here.

	\section{ A parity oblivious communication game }\label{sec:POCG}
We consider a specific communication game in the prepare-measure scenario, which has a structural resemblance with the games in \cite{pan19,lsw}. The game involves two parties, Alice and Bob, who own their respective instruments. Alice receives input $x \in \{1,2,3\}$ which is chosen at random from a uniform probability distribution, i.e., $\forall x $ $p(x)=1/3$, based on which her instrument produces an output $a\in \{0,1\}$. Alice encodes the information about her input $x$ and output $a$ and sends it to Bob. Explicitly, the six different input combinations of Alice are $x^{i}\in (xa)\equiv \{10,11,20,21,30,31\}$, with $i=(1,2,3,..6)$. On the other hand, Bob receives input $y \in \{1,2,3\}$ which is also chosen at random, i.e., $\forall y $ $p(y)=1/3$ and produces an output $b$. The winning condition of the game is that Bob's instrument has to produce the output $b$ such that the following condition 
		\begin{equation}
	  b=\delta_{x,y}\oplus_{2}a,\label{eq:outputform}  
	\end{equation}
 is satisfied. The average success probability of the game is then given by,
	\begin{align}
		\mathcal{A} = \dfrac{1}{18}\sum\limits_{x,y=1}^{3}\sum\limits_{a\in\{0,1\}} p\qty(b=\delta_{x,y}\oplus_{2}a|x, y).\label{eq:sucpro}
	\end{align}
	 Such a game can obviously be won with the probability equal to unity, if Alice is allowed to share unbounded amount of information with Bob. However, in the game, the parity-oblivious constraint limits Alice's communication, such that Alice is restricted from revealing the parity information to Bob. To explain this explicitly, let us consider two  subsets of Alice's input, the even parity set $\mathbb{P}_{e}=\{x\oplus_{2} a=0\}$ and the odd parity  set $\mathbb{P}_{o}=\{x\oplus_{2} a=1\}$. Then $\forall y$, Bob cannot know which parity set the input that belongs to him came from. In an operational theory, the above parity-oblivious restriction  can be formally written as,
\begin{align}
		\forall y,  \ \ \sum\limits_{x^{i}\in \mathbb{P}_{e}} p(b|x^{i},y)=\sum\limits_{x^{i} \in \mathbb{P}_{o}} p(b|x^{i},y).\label{poc}
\end{align}
We derive the maximum classical and quantum success probabilities of the game with the constraints in Eq. (\ref{poc}). 
Generally, classical strategies are described through the ontological models \cite{harrigen, ballentine,banik14,spekkens05} wherein the preparation procedures are characterized by a probability distribution $\mu(\lambda)$ corresponding to an ontic state variable $\lambda$ as discussed earlier.  Following  \cite{spek09}, we argue that the parity-obliviousness at the operational level must also reflect as an equivalent restriction at the level of ontic states if the underlying classical ontological model of the operational theory is preparation non-contextual. Hence, the classical strategy that entails parity-obliviousness at the level of ontic states is essentially a preparation non-contextual strategy.

  Let us now consider that in operational quantum theory,  Alice encodes her input $x^{i}\in \{x,a\}$ into six qubit  states  $\{\rho_{xa}\}$. Then  the parity-obliviousness implies that the following restriction has to be satisfied by Alice's preparation,
		\begin{align}
		\label{poqm}
		\sum\limits_{x^{i}: x\oplus_{2} a=0} \rho_{x^{i}}=	\sum\limits_{x^{i}: x\oplus_{2} a=1} \rho_{x^{i}}
	\end{align}
Explicitly, the parity-oblivious condition reads as $\rho_{11}+\rho_{20}+\rho_{31}=\rho_{10}+\rho_{21}+\rho_{30}$. Hence, if the underlying ontological model of quantum theory is preparation non-contextual, then we must have,

\begin{equation}
\label{po}
\mu(\lambda|P_{11})+\mu(\lambda|P_{20})+\mu(\lambda|P_{31})=\mu(\lambda|P_{10})+\mu(\lambda|P_{21})+\mu(\lambda|P_{30})
\end{equation}
 This parity-obliviousness condition at the ontological level dictates that the classical strategy should be such that any message containing information about the parity of the preparation is forbidden from communication. Imposing the above-mentioned constraint, the optimal success probability of the game in a preparation non-contextual model is derived as $(\mathcal{A})_{pnc}= \frac{13}{18} $. This is explicitly proved  in the following.

In an ontological model the success probability in Eq. \eqref{eq:sucpro} can be written as

\begin{eqnarray}
\label{eq:sucpro1}
	\mathcal{A}=\dfrac{1}{18}\sum\limits_{x,y=1}^{3}\sum\limits_{a\in\{0,1\}}\sum_{\lambda\in\Lambda}\mu(\lambda|P_{xa})\xi(b=\delta_{x,y}\oplus_{2}a|\lambda,M_{y}),
\end{eqnarray}

where $\mu (\lambda|P_{xa})$ is the probability distribution for the preparation of ontic state $\lambda$ given $P_{xa}$ is prepared in the lab, and $\xi(b=\delta_{x,y}\oplus_{2}a|\lambda,M_{y})$ is the response function corresponding to the correct output of Bob's measurement $M_{y}$.

 By expanding Eq. \eqref{eq:sucpro1} and further simplifying it we get,

\begin{eqnarray}
\label{eq:sucp2}
\mathcal{A}=\frac{1}{18}&&\sum_{\lambda}\Bigg[\Big(\mu(\lambda|P_{10})+\mu(\lambda|P_{21})+\mu(\lambda|P_{31})\Big)\xi(b=1|M_{1},\lambda)\nonumber\\
&&+\Big(\mu(\lambda|P_{10})+\mu(\lambda|P_{21})+\mu(\lambda|P_{30})\Big)\xi(b=0|M_{2},\lambda)\nonumber\\
&&+\Big(\mu(\lambda|P_{10})+\mu(\lambda|P_{20})+\mu(\lambda|P_{31})\Big)\xi(b=0|M_{3},\lambda)\nonumber\\
&&+\Big(\mu(\lambda|P_{11})+\mu(\lambda|P_{20})+\mu(\lambda|P_{30})\Big)\xi(b=0|M_{1},\lambda)\nonumber\\
&&+\Big(\mu(\lambda|P_{11})+\mu(\lambda|P_{20})+\mu(\lambda|P_{31})\Big)\xi(b=1|M_{2},\lambda)\nonumber\\
&&+\Big(\mu(\lambda|P_{11})+\mu(\lambda|P_{21})+\mu(\lambda|P_{30})\Big)\xi(b=1|M_{3},\lambda)\Bigg] 
\end{eqnarray}

It is already mentioned that the game is constrained by parity-oblivious restriction given by Eq. (\ref{po}). Applying this constraint relation in Eq. \eqref{eq:sucp2} and simplifying we get
\begin{eqnarray}
\label{eq:sucpf}
\mathcal{A}=\frac{1}{18}\Bigg[9+2&&\sum_{\lambda}\Bigg\{\Big(\mu(\lambda|P_{11})-\mu(\lambda|P_{10})\Big)\xi(b=0|M_{1},\lambda)\nonumber\\
&&+\Big(\mu(\lambda|P_{20})-\mu(\lambda|P_{21})\Big)\xi(b=0|M_{2},\lambda)\nonumber\\
&&+\Big(\mu(\lambda|P_{31})-\mu(\lambda|P_{30})\Big)\xi(b=0|M_{3},\lambda)
\Bigg\}\Bigg] 
\end{eqnarray}
To optimize the above expression we note that the optimization condition requires that the terms having negative signs must be zero, i.e., $\mu(\lambda|P_{10})=\mu(\lambda|P_{21})=\mu(\lambda|P_{30})=0$. But this immediately end up giving the positive terms to be zero, which follows directly from parity-oblivious conditions Eq. \eqref{po} and imposes a trivial bound to the success probability. Thus we can at most set two of the negative terms to be zero such that the parity-oblivious condition now takes the form,
\begin{equation}
\label{newbob}
    \mu(\lambda|P_{30})=\mu(\lambda|P_{11})+\mu(\lambda|P_{20})+\mu(\lambda|P_{31})
\end{equation}
Combining Eq. \eqref{newbob} with Eq. \eqref{eq:sucpf} we can write,
\begin{eqnarray}
\label{eq:sucpff}
\mathcal{A}=\frac{1}{18}\Bigg[9+2&&\sum_{\lambda}\Bigg\{\mu(\lambda|P_{11})\xi(b=0|M_{1},\lambda)\nonumber \\  
&&\hspace{2pt}+\mu(\lambda|P_{20})\xi(b=0|M_{2},\lambda)\\ \nonumber
&&-\Big(\mu(\lambda|P_{11})+\mu(\lambda|P_{20})\Big)\xi(b=0|M_{3},\lambda)\Bigg\}
\Bigg] 
\end{eqnarray}
It is now straightforward to see from Eq. \eqref{eq:sucpff} that a strategy that gives the classical optimal value to the success probability should include ontic state variable $\lambda$ that gives $\sum_{\lambda}\mu(\lambda|P_{11})\xi(b=0|M_{1},\lambda) = \sum_{\lambda}\mu(\lambda|P_{20})\xi(b=0|M_{2},\lambda) =1$, and $\sum_{\lambda}\Big(\mu(\lambda|P_{11})+\mu(\lambda|P_{21})\Big)\xi(b=0|M_{3},\lambda)=0$. Thus the preparation noncontextual classical bound to the success probability of the game becomes $(\mathcal{A})_{pnc}=\frac{13}{18}$. 

A simple strategy can saturate the above bound when Alice makes a trit communication (t) to Bob. For example, suppose Alice sends a trit information $t\in\{1,2,3\}$ in a particular run of experiment that corresponds to preparations from set $\{10,31\}$, $\{11,21\}$ or $\{20,30\}$, respectively. This correspondence is already known to Bob before the game starts.  Now the pre-decided strategy is such that Bob will output $b=0$, $b=1$, and $b=0$ if his inputs are $y=1$, $y=2$, and $y=3$, respectively when trit information $t=1$ is sent. Similarly, for trit information $t=2$, Bob outputs $b=0$, $b=1$, $b=1$ respectively and for trit information $3$, outputs $b=1$, $b=0$, $b=0$ in accordance to the inputs $y=1$, $y=2$, and $y=3$ respectively. Thus in the first and second case, they will win four times among the six runs in each of the cases. Whereas in the third case they win five out of six runs. Thus the success probability becomes $(\mathcal{A})_{pnc}=\frac{13}{18}$, as already derived analytically. It is important to note that the discussed strategy ensures that no information about the parity of the preparations is shared due to Alice's communication of trit information to Bob.

We use the terms 'classical' and 'preparation non-contextual interchangeably throughout the paper. Thus, the classical bound to the success probability will always mean the preparation noncontextual bound derived here.
\section{Sequential quantum advantage in the communication game}\label{sec:SQA}

The sequential sharing of quantum preparation contextuality plays a crucial role in this work in certifying the unsharpness of the measurement instruments. By unsharp measurements, we specifically mean the noisy variants of the sharp measurements so that the number of POVMs is the same as the number of projectors. Although the optimal quantum violation of local or non-contextual bounds facilitates the certification of states and measurements, it does not account for the certification of unsharp POVMs or post-measurement states. The certifications of post-measurement states are inevitable to certify unsharp measurements, which can only be done if a joint sequential quantum violation of the classical bound can be achieved. This, in turn, certifies the states, observables, as well as an unsharpness parameter of the instrument.  
\begin{figure*}
	\centering
	\includegraphics[scale = 0.5]{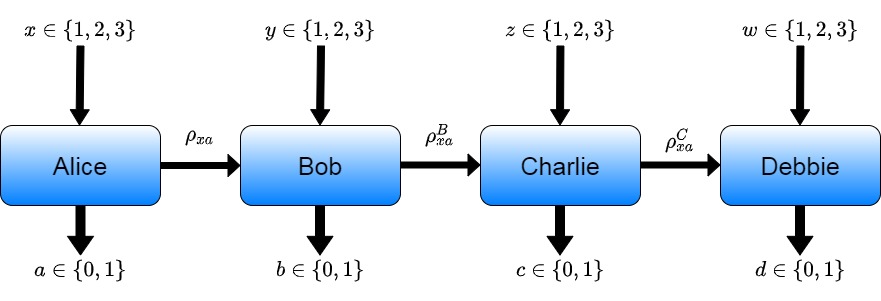}
	\caption{The sequential setting of the game.}
	\label{fig:seqd}
\end{figure*}

As already mentioned, Alice encodes the input $x^{i}\in (x a)$ into qubit states  $\{\rho_{xa}\}$ and sends them to Bob through a unitary channel. Upon receiving the state, Bob performs a  measurement $\{E_{b|y}\}$ depending upon his input $y$ and obtains an output $b$. Now, with the projective measurements, Bob can extract the maximum information from a state discarding all the information about the initial state. In such circumstances, a second observer (say, Charlie) will not be able to extract any information about the initial state. In essence, a posterior observer can obtain no quantum advantage if the prior observer employs a sharp measurement. Therefore, to extend the quantum advantage to Charlie in Fig.\ref{fig:seqd}, Bob must perform the unsharp measurement \cite{busch} characterized by POVMs. In that case, the system will be partially disturbed by keeping residual coherence in the post-measurement state. However, a particular observer can only get the quantum advantage if its unsharpness parameter surpasses a threshold value. This provides the provision of minimal information from the system that is just enough to get the quantum advantage. Hence, whether or not the sequential observer gets the quantum advantage depends on the degree of unsharpness of the preceding observer. Thus, there exists a trade-off between the quantum advantages of sequential observers.

Explicitly, the process runs as follows, as depicted in Fig.\ref{fig:seqd}. Alice prepares qubit states  $\{\rho_{xa}\}$ and sends it to Bob. Subsequently, Bob performs an unsharp measurement with his input $y\in\{1,2,3\}$ and produces an output $b$. Thereby the transformed states are relayed to Charlie, and he performs an unsharp measurement with his input $z\in\{1,2,3\}$ and produces an output $c$. This process goes on for $m$ number of observers $\qty(m \in \{B, C, D...\})$ as long as the $m^{th} $ observer gets the quantum advantage. From the winning condition of the game in Eq.\eqref{eq:outputform} and the success probability in Eq.\eqref{eq:sucpro} we can write the quantum success probability for Bob as,

 \begin{equation}
\label{eq:succB1}
		\mathcal{A}_{B} =\frac{1}{18}\sum_{x,y =1}^3\sum_{a\in \{0,1\}}\Tr\Big[\rho_{xa} E_{b|y}\Big],	
\end{equation}
where $E_{b|y}$ is the POVM corresponding to the outcome $b$ of measurement $B_{y}$.
Now, let us take the prepared state to be,
\ba 
\rho_{xa}= \frac{\mathbb{I}+\va{r}_{xa}\vdot \sigma}{2}, 
\ea 
where $\va{r}_{xa}$ is the Bloch vector and $E_{b|y} = \frac{\I+(-1)^{b}\eta_{B}B_{y}}{2}$ is the POVM corresponding to the output $b$  of measurement of the operator $B_{y}=\hat{b}_{y}.\sigma$ and $ \eta_{B}$ is the unsharpness parameter of Bob's measurement instrument, that quantifies the fuzziness in his measurement. The quantum success probability for Bob can then be written as 
\begin{equation}
	\mathcal{A}_{B}=\frac{1}{2}+\frac{\eta_{B}}{36}\sum_{k,y=1}^3\big(\delta_{k,y}\va{n}_{k}\vdot \vu{b}_{y}\big), \label{eq:succ1}
\end{equation}\\
where $\vu{b}_{y}$ is the unit vector in the direction of the operator $B_{y}$ such that $B_{y}= \vu{b}_{y}\cdot \sigma$, and $\va{n}_{k}=\norm{\va{n}_{k}} \vu{n}_{k}$ ( $ \vu{n}_{k}$ is the unit vector) is the unnormalized Bloch vectors which are explicitly written as,
\begin{align}
	\va{n}_1&=-(\va{r}_{10}-\va{r}_{11})+(\va{r}_{21}-\va{r}_{20})+(\va{r}_{30}-\va{r}_{31}) \nonumber \\
	\va{n}_2&= (\va{r}_{10}-\va{r}_{11})-(\va{r}_{21}-\va{r}_{20})+(\va{r}_{30}-\va{r}_{31}) \nonumber\\ 
	\va{n}_3&= (\va{r}_{10}-\va{r}_{11})+(\va{r}_{21}-\va{r}_{20})-(\va{r}_{30}-\va{r}_{31}). \label{eq:effd}
\end{align}
If Bob's measurement instrument is represented by the set of Kraus operators $\{\mathcal{K}_{{b}|y}\}$ then the reduced state after his measurement can be written as,
\begin{equation}
	\rho_{xa}^{B} = \frac{1}{3}\sum_{y =1}^3\sum_{b\in\{0,1\}}\mathcal{K}_{{b}|y}\rho_{xa}\mathcal{K}_{{b}|y}.  \label{sttr}
\end{equation}
where the Kraus operators satisfy $\sum_{b}\mathcal{K}_{{b}|y}\mathcal{K}_{{b}|y}^\dagger = \mathbb{I}$, with
$\mathcal{K}_{b|y}= \mathcal{U}\sqrt{E_{b|y} }$ for any unitary operator $\mathcal{U}$. As our results are valid for any unitary transformation, for simplicity, we put $\mathcal{U}$=$\mathbb{I}$. The Kraus operators are be written as,	$\mathcal{K}_{\pm|y}= \sqrt{\frac{(1\pm\eta_B )}{2}}\Pi_{+|y} +\sqrt{\frac{(1\mp\eta_B )}{2}}\Pi_{-|y} = \alpha_B  \ \mathbb{I} \pm \beta_B \ B_{y} $ where $\Pi_{\pm|y}=(\mathbb{I}\pm B_{y})/2$ are the projectors corresponding to the qubit observable $B_{y}$, such that,  
\begin{eqnarray}
    \alpha_B &&= \frac{1}{2}\qty(\sqrt{\frac{1-\eta_{B}}{2}}+ \sqrt{\frac{1+\eta_{B}}{2}})\\
    \beta_B &&= \frac{1}{2}\qty(\sqrt{\frac{1+\eta_{B} }{2}}-\sqrt{\frac{1-\eta_{B} }{2}})
\end{eqnarray}
, with $\alpha_B^2  +\beta_B^2 = 1/2$ and $\eta_B= 4\alpha_B\beta_B $.

Using the above form and upon simplifying Eq.\eqref{sttr}  we find  the reduced state after Bob's measurement in its generalized form as,
\begin{equation}
	\rho_{xa}^{B} =\frac{1}{2}\left( \mathbb{I} + \va{r}^{\ B}_{xa}\vdot\sigma\right),\label{st2} 
\end{equation}
where
\begin{equation}
	\va{r}^{\ B}_{xa}= 2\left(\alpha_{B}^2-\beta_{B}^{2}\right)\va{r}_{xa} +\frac{4\beta_{B}^2}{3}\sum_{y=1}^3\left( \vu{b}_{y}\vdot\va{r}_{xa}\right) \vu{b}_{y} \label{eq:vecA}
\end{equation}

is the Bloch vector of the reduced state. Therefore, from Eq.\eqref{eq:sucpro} the quantum success probability for Charlie can be written as,
\begin{equation}
	\begin{split}
		\mathcal{A}_{C} &=\frac{1}{18}\sum_{x,z =1}^3\sum_{a\in\{0,1\}}Tr\Big[\rho_{xa}^{B} E_{c|z}\Big]\\
		&= \frac{1}{2}+\frac{1}{36}\Big[2\left(\alpha_{B}^2-\beta_{B}^{2}\right)\sum_{k,z=1}^3\big(\delta_{k,z}\ \va{n}_{k}\vdot \vu{c}_{z}\big)\\
		&\qquad+\frac{4\beta_{B}^2}{3} \sum_{k,y,z=1}^3\left(\delta_{k,z} \ \va{n}_{k}\vdot \vu{b}_{y}\right)\left( \vu{b}_{y}\vdot \vu{c}_{z}\right)\Big]. \label{eq:succ2}
	\end{split}
\end{equation}
where $E_{c|z} = \frac{\I+(-1)^{c} C_{z}}{2}$ is the POVM corresponding to the result $c=\delta_{x,z}\oplus_{2}a$, of the measurement of $C_{z}=\hat{c}_{z}.\sigma$. The last line comes directly from Eq.\eqref{st2} and Eq.\eqref{eq:vecA}. It is clear from the Eq.\eqref{eq:succ2} that, through its dependence on $\alpha_B$ and $\beta_B$, the success probability $\mathcal{A}_{C}$ of Charlie is a function of $\eta_B$. Hence, there exists a trade off-between $\mathcal{A}_B$ and $\mathcal{A}_C$. We optimize the success probability of Charlie $\mathcal{A}_C$ as a function of $\mathcal{A}_B$.

It is seen from Eq.\eqref{eq:succ2} that as $\alpha_{B}>\beta_{B}$, for the optimization of $\mathcal{A}_{C}$ we must have $\va{n}_{k}$ to be along the direction of $ \vu{c}_{z}$ whenever $k =z$. We further observe that $\va{r}_{xa}$'s have to be antipodal pairs to maximize $\va{n}_{k}$ in Eq.\eqref{eq:effd} such that	$\va{r}_{10}=-\va{r}_{11}$, $\va{r}_{21}=-\va{r}_{20}$ and $\va{r}_{30}=-\va{r}_{31}$. The overall optimization also demands, the states $\{\rho_{xa}\}$ have to be pure i.e., $\norm{\va{r}_{xa}}=1$.  From this choice we can rewrite the $\va{n}_{k}$s as,  $\va{n}_1= 2(-\va{r}_{10}+\va{r}_{20}+\va{r}_{30})$, $\va{n}_2=2(\va{r}_{10}-\va{r}_{20}+\va{r}_{30})$ and 	$\va{n}_3= 2(\va{r}_{10}+\va{r}_{20}-\va{r}_{30})$.  We can then write
\begin{equation}
	\begin{split}
		\mathcal{A}_{C} &\leq  \frac{1}{2}+\frac{1}{36}\Big[2\left(\alpha_{B}^2-\beta_{B}^{2}\right)\sum_{k=1}^3\norm{\va{n}_{k}}\\
		&+\frac{4\beta_{B}^2}{3} \sum_{k, y=1}^3\left(\va{n}_{k}\vdot \vu{b}_{y}\right)\left( \vu{b}_{y}\vdot \vu{c}_{k}\right)\Big].  \label{eq:succ211}
	\end{split}
\end{equation}
Using concavity of the square root $\sum_{k=1}^{3}\norm{\va{n}_{k}} \leq \sqrt{3\sum_{k=1}^{3}\norm{\va{n}_{k}}^{2}}$ and putting the expressions of $\va{n}_{k}$s, we have
\begin{equation}
	\sum_{k=1}^{3}\norm{\va{n}_{k}} \leq 2\sqrt{36-3\left(\va{r}_{10}+\va{r}_{21}+\va{r}_{30}\right)^{2}}. \label{eq:msum}
\end{equation}
Evidently, the maximum value of Eq.\eqref{eq:msum} is given by max$\qty(\sum_{k=1}^{3}\norm{\va{n}_{k}})=12$ when the condition 
\begin{align}
	\label{poco}
	\va{r}_{10}+\va{r}_{21}+\va{r}_{30}=0 
\end{align}
in Eq.\eqref{eq:msum} is satisfied. Using Eqs.\eqref{eq:effd} and \eqref{poco} it is simple to show that the value of each $\norm{\va{n}_{k}}$ is $4$, and hence the equality in Eq.\eqref{eq:msum} holds. The condition in Eq.\eqref{poco} implies the following relations to be satisfied by the Bloch vectors as $\va{r}_{10}\vdot \va{r}_{20}+\va{r}_{10}\vdot \va{r}_{30}=-1$,  $\va{r}_{10}\vdot \va{r}_{20}+\va{r}_{20}\vdot \va{r}_{30}=-1$ and $\va{r}_{10}\vdot \va{r}_{30}+\va{r}_{20}\vdot \va{r}_{30}=-1$. The set of equations in turn provides  $\va{r}_{10}\vdot \va{r}_{20}=\va{r}_{10}\vdot \va{r}_{30}=\va{r}_{20}\vdot \va{r}_{30}$=-1/2.
Impinging the above conditions into Eq.\eqref{eq:effd} we immediately get,   $\vu{n}_{1}=-\va{r}_{10} $, $\vu{n}_{2}=-\va{r}_{20} $ and $\vu{n}_{3}=-\va{r}_{30}$ . 

Without any loss of generality we can then fix a set of vectors that satisfies the above conditions as $\va{r}_{10}=\vu{x}$, $\va{r}_{20}=-\frac{1}{2}\vu{x}+\sqt\vu{z}$ and $\va{r}_{30}=-\frac{1}{2}\vu{x}-\sqt\vu{z}$. Furthermore, to optimize the quantum success probability of Charlie $\mathcal{A}_{C}$ of Eq.\eqref{eq:succ2} we also get $\va{n}_{k}=\va{c}_{z}$ for $z=k$. This implies that the maximum value $max\left(\mathcal{A}_{C}\right)\equiv\Omega_{C}$ can be obtained when the unit vectors of Charlie are $ \vu{c}_{1}=\vu{b}_{1}=\vu{n}_{1}$, $ \vu{c}_{2}=\vu{b}_{2}=\vu{n}_{2}$ and $ \vu{c}_{3}=\vu{b}_{3}=\vu{n}_{3}$.   We then have,
\begin{equation}
	\label{eq:succ2d}
	\mathcal{A}_{C}\leq\frac{1}{2}+\frac{\left(3\alpha_{B}^2-\beta_{B}^2\right)}{54} \sum\limits_{k=1}^3\norm{\va{n}_{k}}.
\end{equation}
It can be easily seen that given the values of $\alpha_{B}$ and $\beta_{B}$, the maximization of $\mathcal{A}_{C}$ provides the success probability of Bob of the form,
\begin{equation}
	\mathcal{A}_{B}=\frac{1}{2}+\frac{\eta_B}{36}\sum_{k=1}^3\norm{\va{n}_{k}}\label{eq:succ12}.
\end{equation}
Note that both $\mathcal{A}_{B}$ and $\mathcal{A}_{C}$ are simultaneously optimized when the quantity $\sum_{k=1}^3\norm{\va{n}_{k}}$ is optimized. Let us denote max$\qty(\mathcal{A}_{B})$ as $\Omega_{B}$. Substituting  max$\qty(\sum_{k=1}^{3}\norm{\va{n}_{k}})=12$, we thus get the optimal pair of success probabilities for Bob and Charlie as, 	
\ba
\label{omegab}
	\Omega_{B} &=& \frac{1}{2}\left(1+\frac{2\eta_{B}}{3}\right).
\ea
\ba
\label{omegac}
 \Omega_{C} &=& \frac{1}{2}\left(1+ \frac{2\eta_{C}(1+2\sqrt{1-\eta_{B}^2})}{9}\right). 
\ea 

\begin{figure}[ht]
\label{fig2}
	\centering
	\includegraphics[scale = 0.65]{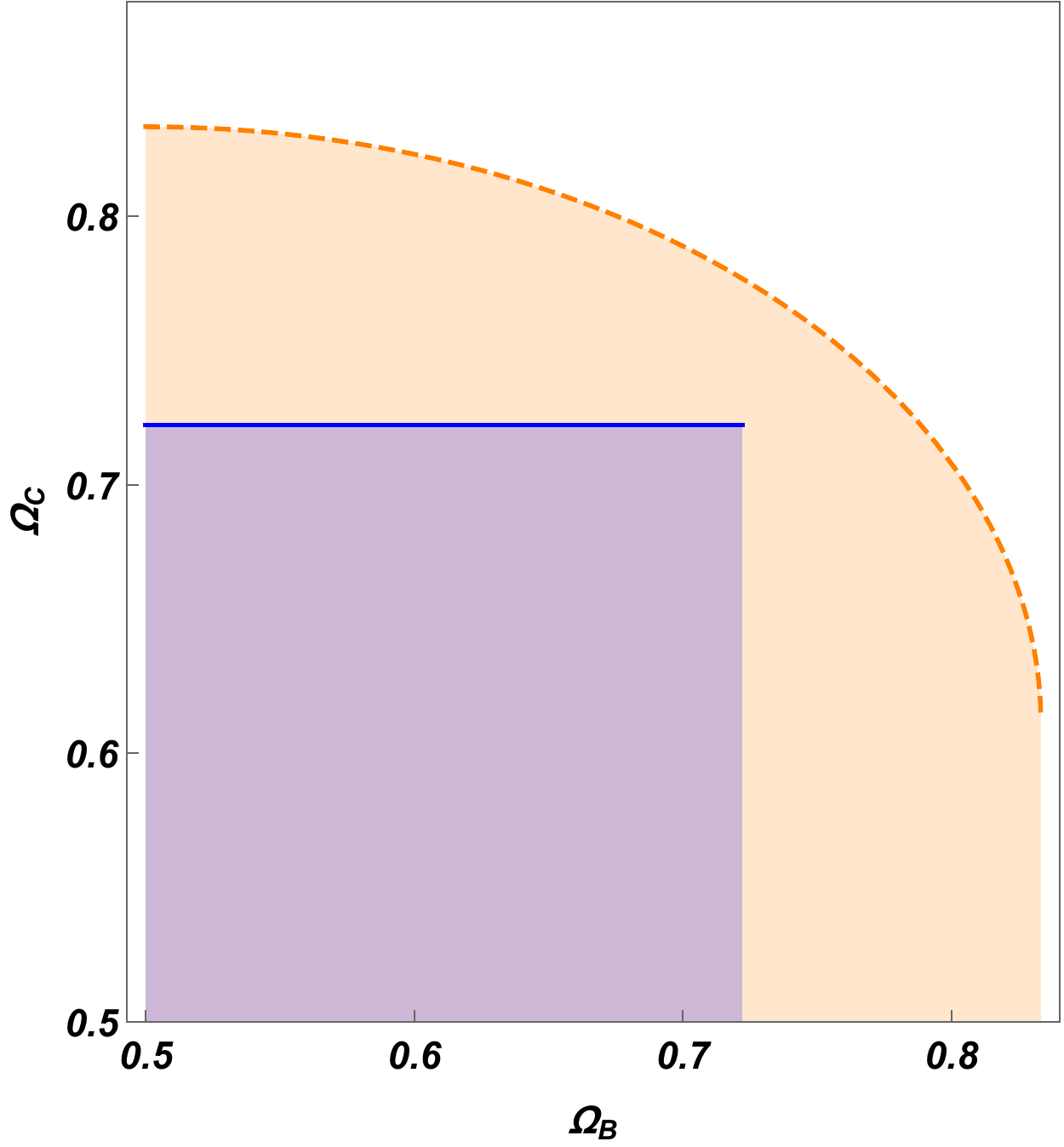}
	\caption{Trade-off relation between the success probabilities achieved by Bob($\Omega_B$) and Charlie($\Omega_C$). The solid blue (dotted orange) line indicates the variation in success probability of Charlie with respect to Bob in classical(quantum) regimes. It is clear that, by classical means there exist no trade-off as there exist no variation in Charlie's success probability with change in Bob's success probability, whereas in quantum mechanical prescriptions there is a solid trade-off in success probabilities between Charlie and Bob.}\label{tro}
\end{figure}
To certify the unsharpness parameter $\eta_B$, we assume Charlie performs a sharp measurement (with $\eta_{C} =1$). In such a case, both $\Omega_{B}$ and $\Omega_{C}$ become functions of a single parameter $\eta_{B}$. 

It is worth pointing out that the set of states which provides the optimal pair of success probabilities($\Omega_{B},\Omega_{C}$) must also comply with the parity-oblivious restriction \eqref{poqm}. In the current scenario this means that the  quantum states corresponding to the set $\mathbb{P}_{e} \in \{10,20,30\}$ and the set $\mathbb{P}_{o} \in \{11,21,31\}$  must produce statistics that are equivalent to Bob for all measurements. Explicitly, Alice's input states must satisfy the following relation, $\rho_{10}+\rho_{20}+\rho_{30}=\rho_{11}+\rho_{21}+\rho_{31}$. One can easily verify that the set of states that are self-tested through the optimal quantum advantage follows the above relational constraint.

\subsection{Certification of the unsharpness parameter }\label{sec:CoL}
From Eq.\eqref{omegac} we note that $\Omega_{C}$ is a function of $\Omega_{B}$. The trade-off relation between the optimal pair $(\Omega_{B},\Omega_{C})$ quantum success probabilities for Bob and Charlie can then be written as,
\begin{equation}\label{omega1to2}
	{\Omega_{C}({\Omega_{B}})= \frac{1}{2}+\frac{\left(1+\sqrt{4-9(2\Omega_{B}-1)^2}\right)}{9}},
\end{equation}
In Fig.\ref{tro} we plot the trade-off between the success probabilities $\Omega_{C}$ and ${\Omega_{B}}$ described by classical and quantum theories. Classically, the measurement does not necessarily disturb a system, and there exists no trade-off between the success probabilities, which is represented by the blue line as shown in Fig.\ref{tro}. On the other hand, due to the information disturbance relations $\Omega_B$ and $\Omega_C$ follow a trade-off relation in quantum theory description as expressed in the Eq.\eqref{omega1to2}. Such trade-off between the quantum success probabilities is captured in Fig.\ref{tro} by the red curve, and each point on it certifies a unique value of unsharpness parameter $\eta_{B}$. For instance, when $\Omega_{B}=\Omega_{C}= 0.75457$, the value of $\eta_{B}= 0.7637$ is certified. A similar argument holds for each point in the red curve in Fig.\ref{tro}.

It is quite interesting to check the extent to which the value of the unsharpness parameter for observers can be fine-tuned to find the maximum number of independent observers that can sequentially harness the quantum advantage. The sequential observers mutually agree that each of them gets an appropriate amount of quantum advantage in the game by setting up the unsharpness parameters of their measurement instruments. Bob's measurement instrument should have its unsharpness parameter above a certain threshold value. Otherwise, he will not receive any quantum advantage in such a game. By using Eq.\eqref{omegab} the minimum value $(\eta_{B})^{min}$ of unsharpness parameter required to get the quantum advantage for Bob is determined as a function of $(\Omega_B)$ as,

\begin{equation}
   \eta_B \geq (\eta_{B})^{min}= 3\left(\Omega_B-\half\right) \label{eq:minl}
\end{equation}
Similarly, to upper bound the permissible value for $\eta_B$, from Eq.\eqref{omegac} we rewrite $\eta_B$ in terms of $\Omega_C$ as, 
\begin{equation}
    \eta_B \leq \qty(\eta_{B})^{max}= \sqrt{1-\qty(\half\qty(\frac{9}{2}\qty(2\Omega_C-1)-1))^2} \label{eq:maxl}
\end{equation}
From Eq.\eqref{eq:minl} and Eq.\eqref{eq:maxl} by substituting the classical bound $((\mathcal{A})_{pnc} = 13/18)$ for $\Omega_B$ and $\Omega_C$, we find the quantitative bound to the unsharpness parameter $\eta_{B}$ of Bob's measurement instrument as,
\begin{equation*}
    \frac{2}{3} \leq \eta_B \leq \frac{\sqrt{3}}{2}
\end{equation*}
Certainly, any value of $\eta_{B}$ in the above range necessarily provides the quantum advantage to Both the observers.
\subsection{SDI certification statements}\label{sec:IST}
The optimal pair $(\Omega_B, \Omega_C)$ of success probabilities for Bob and Charlie provides the following certification statements for the ideal preparations of Alice and the ideal unsharp measurements of Bob.

(i) Among the six states that Alice prepares, three are of a specific type called the trine states, and the rest are just their antipodals. For an example, such states are represented by the Bloch vectors, $\va{r}_{10}=\vu{x}$, 	$\va{r}_{20}=\qty(\frac{1}{2}\vu{x}-\frac{\sqrt{3}}{2}\vu{z})$, $\va{r}_{30}=\qty(-\frac{1}{2}\vu{x}-\frac{\sqrt{3}}{2}\vu{z})$ and their respective antipodal pairs.

(ii) Bob performs unsharp measurements corresponding to the incompatible observables such that, $\vu{b}_{1}=-\va{r}_{10}$, $\vu{b}_{2}=-\va{r}_{20}$ and $\vu{b}_{3}=-\va{r}_{30}$. Charlie's measurement settings are the same as Bob's with $\eta_{C}$ instead of $\eta_B$ and are taken to be unity as Charlie performs a sharp measurement.

Furthermore, each point on red curve in Fig.\ref{tro} is a representative of the optimal pair $(\Omega_{B},\Omega_{C})$ for a certain $\eta_{B}$. The certification arguments are valid as long as for a certain $\eta_{B}$ the observers get optimal pairs. In other words, every point on the red curve in Fig.\ref{tro} represents a certified value of $\eta_B$. It can be easily seen from Fig.\ref{tro} that whether or not Charlie gets quantum advantage entirely depends upon how much of the same is already extracted from the system by Bob's measurement. They can then agree on a common strategy to achieve the required quantum advantage for a particular purpose.

\section{Possible extension of quantum advantage to a third  observer}\label{sec:Debbie}

We have just demonstrated that in a sequential scenario, a sustainable quantum advantage is achieved by Charlie if Bob performs unsharp measurements for a range of values of $\eta_{B}$. A question may immediately arise that for sufficiently lower values of $\eta_{C}$ whether it is possible to can be extended the quantum advantage to a third observer (say, Debbie). To analyze this case, we find the maximum success probability for Debbie, who receives input $w\in \{1,2,3\}$ and returns output $d\in \{0,1\}$. By using  Eq.\eqref{sttr} we find the reduced states after Charlie's measurement, relayed to Debbie as,
\begin{align}
	\rho_{xa}^{ C} = \half\left(\mathbb{I}+\va{r}_{xa}^{\ C}\vdot\sigma\right)
\end{align}
where
\begin{eqnarray}
\va{r}_{xa}^{\ C}&&= 4\left(\alpha_{B}^2-\beta_{B}^{2}\right)\left(\alpha_{C}^2-\beta_{C}^{2}\right)\va{r}_{xa}+\frac{8}{3}\beta_{B}^{2}\left(\alpha_{C}^2-\beta_{C}^{2}\right) \sum_{y=1}^3\left( \vu{b}_y\vdot\va{r}_{xa}\right) \ \vu{b}_y \nonumber \\ &&\qquad+\frac{8}{3}\ \beta_{C}^{2}\left(\alpha_{B}^2-\beta_{B}^{2}\right) \sum_{z=1}^3\left( \vu{c}_z\vdot\va{r}_{xa}\right) \ \vu{c}_z\nonumber\\ \nonumber &&\qquad+\frac{16\ \beta_{B}^2\ \beta_{C}^{2}}{9}\sum_{y,z=1}^3\left( \vu{b}_y\vdot\va{r}_{xa}\right)\left( \vu{b}_y\vdot \vu{c}_z\right) \ \vu{c}_z.
	\end{eqnarray}
Then the quantum success probability for Debbie is given by,
\begin{equation}
\begin{split}
 \mathcal{A}_{\ D} &= \frac{1}{18}\Tr\left[\rho_{xa}^{C}E_{d|w}\right]\\
&= \half +\frac{1}{36}\Bigg[4\qty(\alpha_{B}^2-\beta_{B}^2)\ \qty(\alpha_{C}^2-\beta_{C}^2)\sum_{k,w,z=1}^{3} \ \delta_{k,w} \ \va{n}_k\vdot \vu{c}_{z}\\
&\qquad+\frac{8}{3}\ \beta_{B}^2\qty(\alpha_{B}^2-\beta_{B}^2)\ \sum_{k,y,w=1}^{3} \qty(\vu{b}_{y} \vdot \va{n}_k)\ \qty(\vu{b}_{y}\vdot \vu{d}_{w})\\
&\qquad+ \frac{8}{3}\ \beta_{C}^2\qty(\alpha_{B}^2-\beta_{B}^2)\sum_{k,z,w=1}^{3}  \qty(\vu{c}_z \vdot \va{n}_k)\ \qty(\vu{c}_z\vdot \vu{d}_{w})\\&\qquad+\frac{16}{9}\ \beta_{B}^2\ \beta_{C}^2\sum_{k,w,y,z=1}^{3}  \qty(\vu{b}_{y} \vdot \va{n}_k) \ \qty(\vu{b}_{y}\vdot \vu{c}_z)\ \qty(\vu{c}_{z}\vdot \vu{d}_w)\Bigg],\label{succD}
\end{split}
\end{equation}
where $E_{d|w} = \frac{\mathbb{I}+(-1)^{d} \eta_{D}D_{w}}{2}$ is the POVM corresponding to the outcome  $d=\delta_{x,w}\oplus_{2}a$ of measurement $D_{w}=\vu{d}_w\vdot\sigma$. Using the same optimization technique used in the previous section from  Eq.\eqref{succD} we obtain,
\begin{equation}
\begin{split}
\mathcal{A}_{\ D} &= \half +\frac{1}{36}\Bigg[4\qty(\alpha_{B}^2-\beta_{B}^2)\ \qty(\alpha_{C}^2-\beta_{C}^2)\sum_{k=1}^{3} ||\va{n}_k||^2\\ & \qquad+\frac{8}{3}\ \beta_{B}^2\ \qty(\alpha_{B}^2-\beta_{B}^2)\sum_{k=1}^{3} ||\va{n}_k||^2\\
&\qquad+ \frac{8}{3} \ \beta_{C}^2\ \qty(\alpha_{B}^2-\beta_{B}^2)\sum_{k=1}^{3} ||\va{n}_k||^2 + \frac{16}{9}\ \beta_{B}^2\ \beta_{C}^2\sum_{k=1}^{3} ||\va{n}_k||^2 \Bigg] \\
\end{split}
\end{equation}
Further by putting $\sum_{k=1}^{3} ||\va{n}_k||^2 = 12$, we finally get, 
\begin{equation}
    \Omega_{D}=\half\left(1+\frac{2\eta_{D}}{27}\left(1+2\sqrt{1-\eta_{B}^2}\right)\left(1+2\sqrt{1-\eta_{C}^2}\right)\right)\label{eq:succk}
\end{equation}
Thus, the unsharpness parameter $\eta_{D}$ of Debbie's measurement instrument takes the following form,
\begin{equation}\label{eq:l3}
    \begin{split}
       \eta_{D} &= \frac{27}{2}\left(\frac{2\Omega_D-1}{\left(1+2\sqrt{1-\eta_B^2}\right)\left(1+2\sqrt{1-\eta_{C}^2}\right)}\right).
    \end{split}
\end{equation}

From the equation above, it is clear that, for Debbie to achieve quantum advantage, the value of $\eta_{C}$ and $\eta_B$ should be such that $\eta_{D}\leq 1$. The range of $\eta_B$ has been specified earlier. Let the threshold value of the unsharpness parameter of Charlie's measurement instrument be denoted by $ (\eta_{C})^{min}$. From Eq.\eqref{omegac} we write,
\begin{equation}
\begin{split}
   (\eta_{C})^{min}&=\frac{9}{2}\left(\frac{2\Omega_C - 1}{1+2\sqrt{1-\eta_{B}^{2}}}\right) \label{lamdaR}
\end{split}
\end{equation}
To give a quantum advantage to Charlie, Bob performs his measurement by setting the unsharpness parameter of his instrument slightly above the lower critical value $(\eta_{B})^{min}$. We denote the infinitesimal amount of $\eta_{B}$ that is greater than $(\eta_{B})^{min}$ by $\zeta\in[0,1]$ such that, $0.66\leq\eta_B + \zeta\leq 0.866$. 

By putting $\eta_{B}=\eta_{B}+\zeta$ in Eq.\eqref{lamdaR} and upon solving we get $(\eta_{C})^{min} = 0.803 + 0.577\zeta$ with $\zeta \in [0,0.341]$. Therefore, lowering the degree of unsharpness parameter of Charlie may provide some advantage. To certify that Debbie can harness the quantum advantage, from Eq.\eqref{eq:l3}, we find the minimum value of $\eta_{D}$ to violate the non-contextual bound. For that, we put $\eta_{B}=(\eta_{B})^{min}$ and $\eta_{C}=(\eta_{C})^{min}$ in Eq.\eqref{eq:succk}. This gives $\eta_{D}= 1.09 $ which is not a legitimate value of the unsharpness parameter. This suggests that irrespective of how small the first two observers impose the disturbances, not more than two observers can get simultaneous quantum advantages.

\section{Robust Certification of preparations and measurements }\label{sec:RST}
In Sec.\ref{sec:IST}, we discussed the certification of trine states, measurements and the unsharpness parameter in an ideal scenario. The practical scenario, however, is bound to be imperfect and hence demands a certification scheme tolerant to noise. We now analyze the robustness of preparation and measurement instruments in the presence of noise. For this, we adopt a scheme proposed in \cite{kaniewski}, and subsequently used in SDI-scenario in \cite{tava2018}. In \cite{kaniewski} the author derived an operator inequality for the DI self-testing bound on the quantum value Clauser-Horne-Shimony-Holt expression by analyzing the robustness of the instruments. Following that approach, we construct such operator inequalities to characterize the average fidelity. Relying on the value of average fidelity, the closeness between the target states (measurements) to ideal ones can be determined. Hence obtaining a lower bound to the average fidelity can provide the robust certification. 

\subsection{Robust certification of the preparations of Alice}
Given a set of state preparations \{$\rho_{xa}$\}, the average fidelity for a given success probability $\mathcal{A}_{B}$ with respect to the set of ideal states $\{\rho_{xa}^{\ ideal}\}$ is quantified by   
\ba
\mathcal{S}(\rho_{xa})=\frac{1}{6}\underset{\Lambda}{max}\sum_{x=1}^3\sum_{a\in\{0,1\}} F\qty(\rho_{xa}^{\ ideal}, \Lambda(\rho_{xa})),
\ea
 where $\Lambda$ is a completely positive trace preserving(CPTP) map and fidelity takes the following form,
 \ba
 F\qty(\rho_{xa}^{\ ideal}, \Lambda[\rho_{xa}])=\braket{\Lambda\left[\rho_{xa}\right]}{\rho_{xa}^{\ ideal}}.\ea
There may exist multiple set of states which are compatible with a given  success probability $\mathcal{A}_{B}$. Thus, we estimate the robustness of the preparations  by finding the lower bound of the average fidelity for all possible set of states that provides the success probability $\mathcal{A}_{B}$ such that,
 \begin{figure}
 	\centering
 	{{\includegraphics[width = 8.5cm]{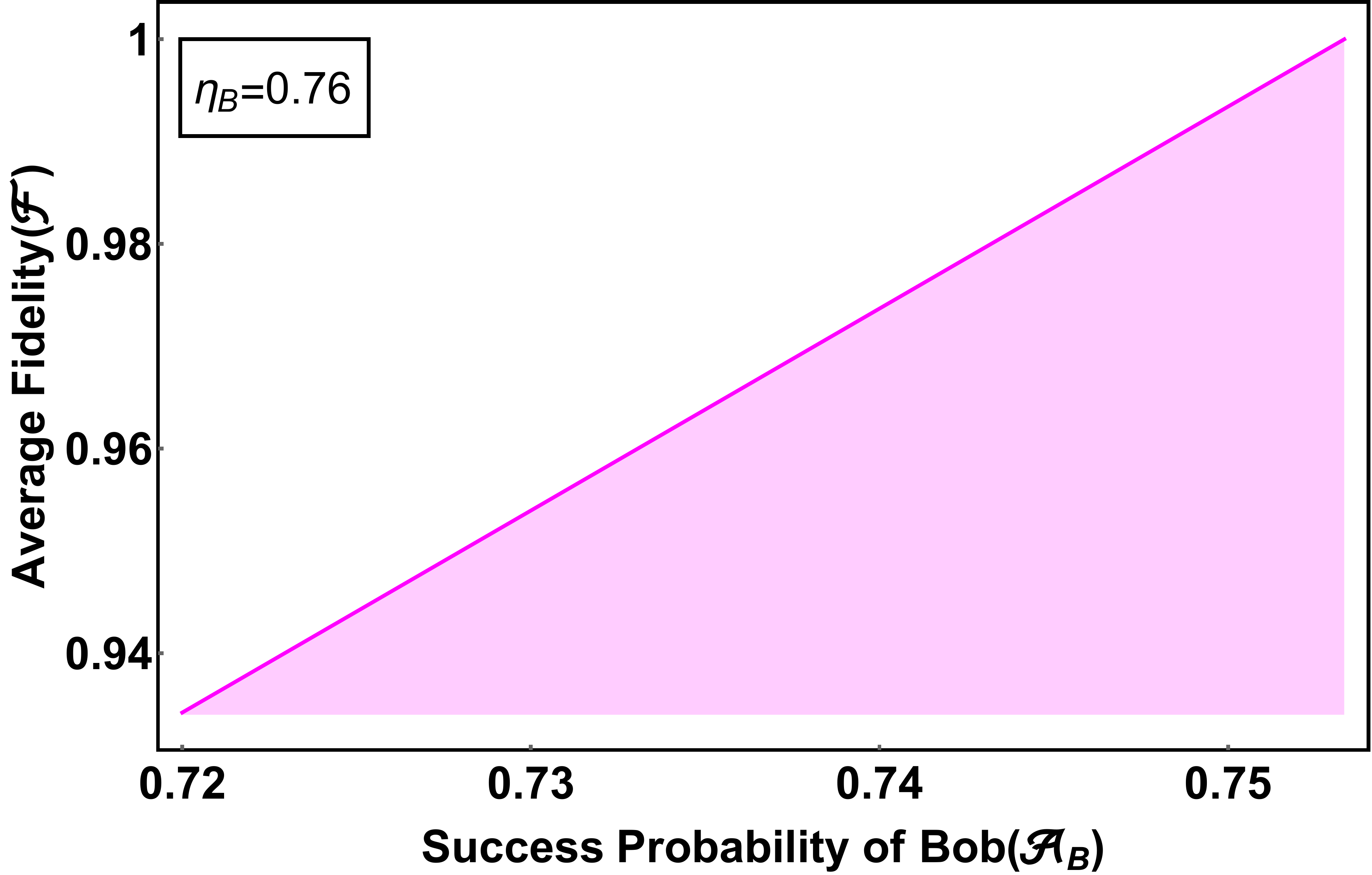}}}\\
 	\caption{ The plot describes the variation between the average fidelity ($\mathcal{F}$) for prepared states of Alice and the success probability ($\mathcal{A}_{B}$)  for a fixed value of unsharpness parameter $\eta_B$.}
 	\label{fig:robustnessp}
 \end{figure}
\begin{align}
	\mathcal{F}(\mathcal{A}) = \underset{\rho_{xa}\in R(\mathcal{A}_{B})}{min} \mathcal{S}({\rho_{xa}}).
\end{align}
 To ensure the robustness of Alice's preparations, the fidelity must be lower bounded by a function of $\mathcal{A}_{B}$ in order to satisfy an appropriate operator inequality. By using Eq.\eqref{eq:succB1} we re-write the success probability of Alice in the following form as,
  \ba \mathcal{A}_B=\half+\sum\limits_{x=1}^3 \sum\limits_{a\in\{0,1\}}Tr[\rho_{xa}W_{xa}]\ea, where
\begin{equation}
     W_{xa} = \frac{1}{36}\sum_{y=1}^3\sum_{a\in\{0,1\}}(-1)^{\delta_{x,y}\oplus a} B_y.
\end{equation}

 If the preparations are pure then the fidelity can be written as $F\left(\rho_{xa}^{\ ideal},\Lambda\left[\rho_{xa}\right]\right) = \braket{\Lambda\left[\rho_{xa}\right]}{\rho_{xa}^{\ ideal}} $. We can then write $F\left(\rho_{xa}^{\ ideal},\Lambda\left[\rho_{xa}\right]\right) = \Tr\qty(K_{xa}\rho_{xa}) $, where
\begin{align}
	K_{xa}(B_1,B_2,B_3) = \Lambda^\dagger(B_1,B_2,B_3)\left[\rho_{xa}^{\ ideal}\right],
\end{align}
represents the action on the set of ideal states with $\Lambda^{\dagger}$ being dual channel to $\Lambda$. For any positive quantity $s$ we define an operator in the form $K_{xa}-sW_{xa}$ and let $t_{xa} \in \mathbb{R}$ be its lower bound. Then, in general the operator inequality takes the form,
\begin{align}\label{opineq}
	K_{xa}(B_1,B_2,B_3)\geq sW_{xa}+t_{xa}(B_1,B_2,B_3)\mathbb{I}
\end{align}
 Finding the trace over the inputs of this inequality and applying minimization over $B_1$, $B_2$ and $B_3$, the lower bound to the average fidelity turns out to be
\ba
\frac{1}{6}\sum_{x=1}^3 \sum_{a\in\{0,1\}} Tr[K_{xa}\rho_{xa}]\geq \frac{s}{6}Tr[\rho_{xa}W_{xa}+t_{xa}\mathbb{I}]
\ea
so that
\ba
\mathcal{S}\geq  \ \frac{s}{6}[(\mathcal{A}_B-\frac{1}{2})]+\frac{1}{6}\sum_{x=1}^3 \sum_{a\in\{0,1\}}t_{xa}.
\ea
We thus bound the average fidelity from below as,
\begin{align}\label{eq:fidelityp}
	\mathcal{F}(\mathcal{A}_B)\geq
	\frac{s}{6}(\mathcal{A}_B-\frac{1}{2}) + t,
\end{align}
where $$t=\underset{B_1,B_2,B_3}{min} \frac{1}{6}\sum_{x=1}^3 \sum_{a\in\{0,1\}}t_{xa}(B_1,B_2,B_3)$$, are subject to evaluation. 

Here, we propose a generalized scheme where robustness can be certified for any choice of the unsharpness parameter$(\eta_B)$ in the defined range for Bob's measurement instrument. The detailed derivation is lengthy and hence  deferred for the Appendix \ref{App:A}. We first fix $s = 9/\eta_B$, which in turn gives $t = \half$ such that, the lower bound of average fidelity can be determined from Eq.\eqref{eq:fidelityp} for a given success probability $\mathcal{A}_B$ as, 
\begin{equation}
    \mathcal{F}\qty(\mathcal{A}_B)=\frac{3}{2\eta_B}\qty(\mathcal{A}_B-\half)+\half.
\end{equation}
The trade-off relation between the average fidelity and the success probability is plotted for $(\mathcal{A})_{pnc}\leq\mathcal{A}_B\leq\Omega_B$ in Fig. \ref{fig:robustnessp} for a particular value of $\eta_{B} = 0.76$.  Due to the dependence of $s$ on $\eta_{B}$,  although the nature of the trade-off graph will be the same for different values of $\eta_{B}$, the fidelity for different degrees of unsharpness will be different even for the same value of success probabilities. The lower bound saturates with $F(\mathcal{A}_B) = 1$ for $\mathcal{A}_B = \Omega_B$.

 \subsection{Robust certification of measurements}
Analogous to robust certification of preparations, we adopt the similar treatment here.
The expression of average fidelity, given the set of ideal measurements is given by,
\ba 
\mathcal{S}'(\{E_{b|y}\}) = \frac{1}{6}\underset{\Lambda}{max}\sum_{y=1}^3\sum_{b\in\{0,1\}}F\qty((E_{b|y})^{\ ideal},\Lambda[E_{b|y}])
\ea
where $\Lambda$ is a quantum channel. To estimate the lower bound for the average fidelity, we write,

\begin{figure}
 	\centering
 	\includegraphics[width = 8.5cm]{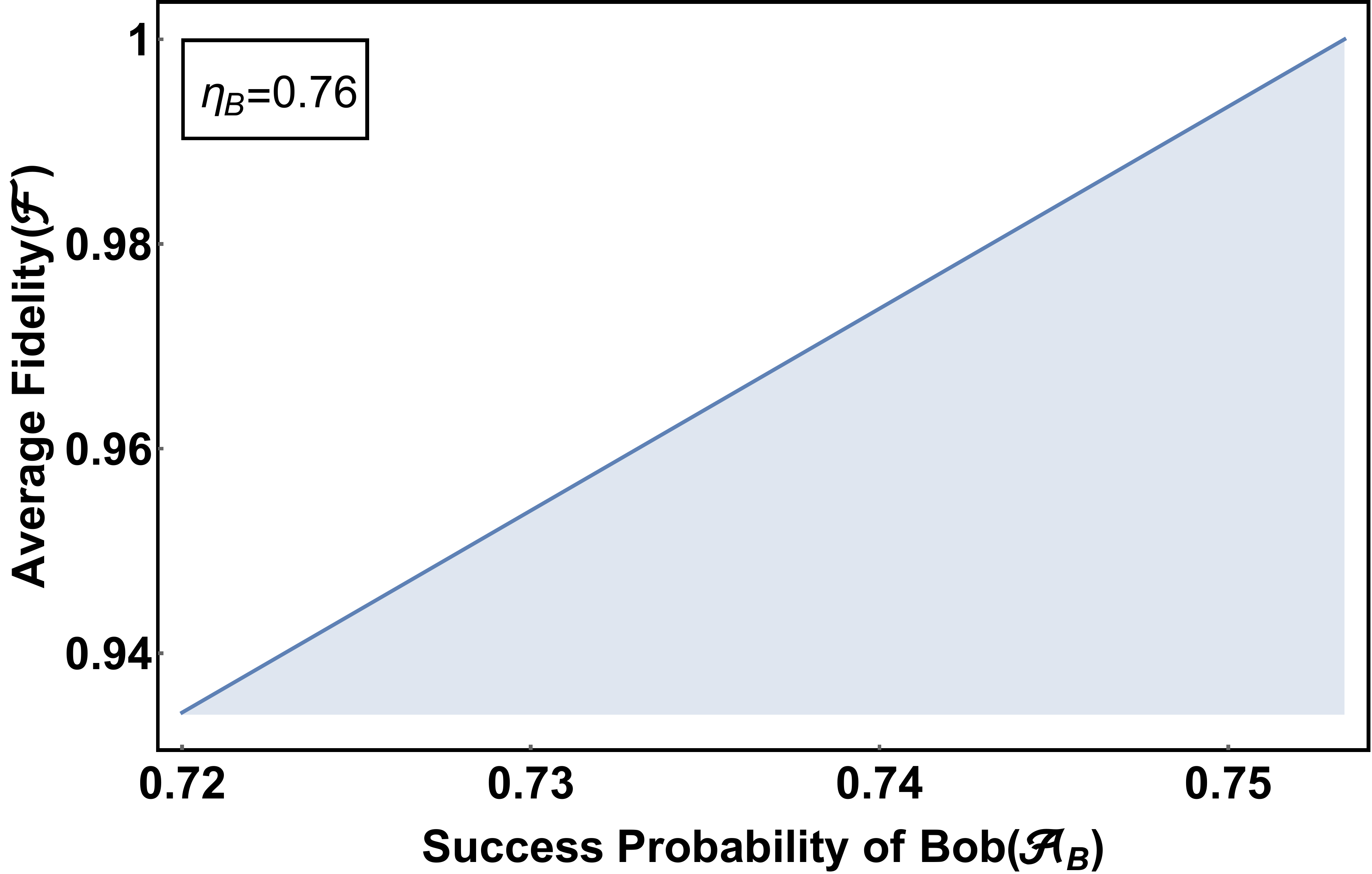}\\
 	\caption{ The plot represents the variation in the average fidelity of measurements of Bob with respect to success probability $\mathcal{A}_B$ for a fixed value of $\eta_B$.
 	}
 	\label{fig:robustnessmb}
 \end{figure}
\begin{align}
	\mathcal{F}'(\mathcal{A}) = \underset{{E_{b|y}}\in R'(\mathcal{A})}{min}\mathcal{S}'(\{E_{b|y}\}).
\end{align}
Measurements are suitably chosen to be compatible with the given value of $\mathcal{A}$ because the instruments are considered to be black boxes. To determine the robustness the operator inequality is constructed of the form
\begin{align}
	K_{yb}(\{\rho_{yb}\}) \geq sZ_{yb}+t_{yb}\mathbb{I}, \label{oe}
\end{align}
where,  $K_{yb}=\Lambda^\dagger[(E_{b|y})^{\ ideal}]$ and thus fidelity can be  written as, 
\ba 
F\qty((E_{b|y})^{\ ideal},\Lambda[E_{b|y}]) &=& \braket{\Lambda\left[E_{b|y}\right]}{(E_{b|y})^{\ ideal}} \\
\nonumber
&=&\Tr\qty[K_{yb}(E_{b|y})].
\ea
From the definition of general success probability $\mathcal{A}=\sum_{y=1}^3\sum_{b\in\{0,1\}}Tr[E_{b|y}Z_{yb}]$, where
\ba 
Z_{yb}=\frac{1}{18}\sum_{x,y=1}^3\sum_{a\in\{0,1\}}\rho_{xa}\delta_{l,b}.
\ea
with $l=\delta_{x,y}\oplus a$. 
 Subsequently, we  find the expression which quantifies the lower bound of the average fidelity to be
\begin{align}
\label{eq:fidelitym}
	\mathcal{F}'(\mathcal{A}) \geq \frac{s}{6}\mathcal{A} + t,
\end{align}
 
where 
\begin{equation}
    t=\frac{1}{6}\text{min}_{B_1,B_2,B_3}\sum_{y=1}^3\sum_{b=\{0,1\}}t_{yb}(B_1,B_2,B_3).
\end{equation}
 Furthermore, the states that Bob receives are the states sent by Alice and the performed measurements are unsharp. On the other hand, Charlie gets the $\eta_{B}$ dependent reduced states after Bob's measurement and performs sharp measurements. As a result, the robustness analysis for these two observers differs which are the following.\\

\textbf{\underline{\textit{For Bob:}}} Bob performs unsharp measurements with unsharpness parameter $\eta_B$ in the certified range. Thus, essentially the parameter $t$ becomes a function of $\eta_{B}$ which finally end up giving the $\eta_{B}$-dependence of the fidelity equation. To lower bound the average fidelity, we choose $s= 9$, and by simple calculation we find $t= \qua-\frac{\eta_B}{2}$. From Eq.\eqref{eq:fidelitym} we can then write ,
\begin{equation}
    \mathcal{F'}\qty(\mathcal{A}_B) \geq \frac{3}{2}\mathcal{A}_B+\qua-\frac{\eta_B}{2}.\label{eq:fidmb}
\end{equation}
For certain $\eta_{B}$ we recover the ideal measurements that we aimed to certify, when the success probability $\mathcal{A}_B$ reaches its maximum value i.e., $\mathcal{F'}\qty(\mathcal{A}_B)=1$ when $\mathcal{A}_B = \Omega_B$ .\\

\textbf{\underline{\textit{For Charlie:}}} Even though Charlie's measurement is sharp, still $t$ is dependent on $\eta_B$ due to the $\eta_B$ dependency in the post measurement state by a factor $\gamma = \qty(1+\sqrt{1-\eta_B^2})/2$. Here we fix $s$ = $54/(7\gamma -1)$,  and upon further simplification the value for $t$ turns out to be, $t=(9-6\gamma)/(2-14\gamma)$. Hence, the lower bound  to the average fidelity is written from Eq.\eqref{eq:fidelitym} as,
\begin{equation}
    \mathcal{F'}\qty(\mathcal{A}_C) =\frac{9}{7\gamma-1}\mathcal{A}_C + \frac{(9 - 6 \gamma)}{(2 - 14 \gamma)}. \label{eq:cff}
\end{equation}

For a certain value of $\eta_{B}$ in the certified range, the fidelity reaches its maximum value for $\mathcal{A}_C = \Omega_C$.

\begin{figure}
 	\centering
 	{{\includegraphics[width = 8.5cm]{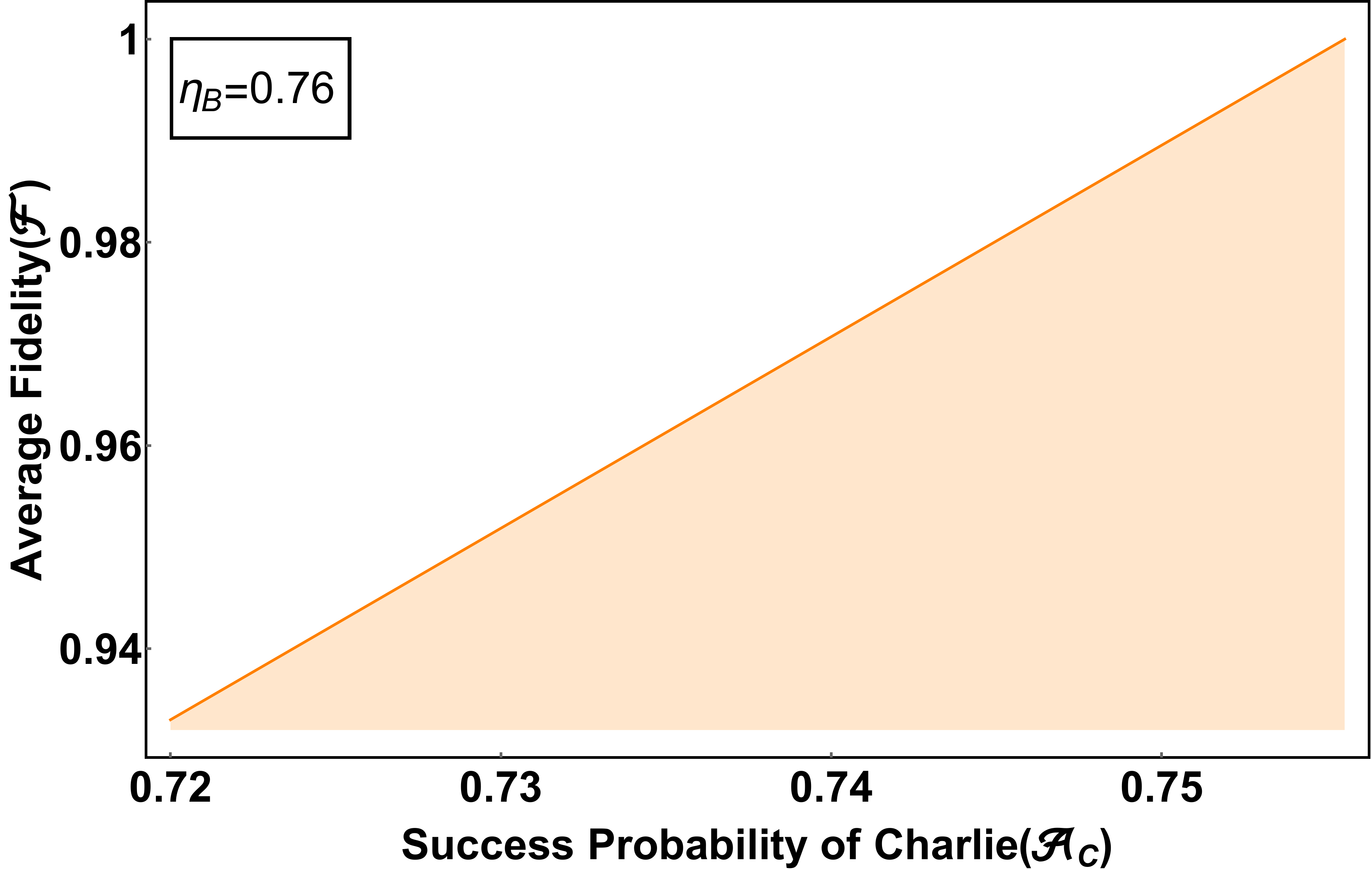}}}\\
 	\caption{ 
 The plot represents the variation in the average fidelity of measurement with the variation of the success probability of Charlie.}
 	\label{fig:robustnessmc}
 \end{figure}

The variation of fidelity with the success probability for Bob and Charlie are plotted in Fig.\ref{fig:robustnessmb} and Fig.\ref{fig:robustnessmc} respectively for a particular choice of $\eta_B$.
Detailed calculation for robust certification of the measurement instrument for Bob and Charlie are provided in the Appendix.\ref{App:B}.\\

\section{Summary and Discussion}\label{sec:SandD}

In summary, we have demonstrated robust certification of states, measurements, and the unsharpness parameter based on the sequential quantum advantage through a specific communication game in a semi-device-independent prepare- measure scenario. We explicitly derived the classical preparation non-contextual bound of the proposed game to quantitatively analyze the quantum supremacy in such a task. The viable constraint relations among the states and observable for the optimal quantum success probability are derived. From these constraints, we further argued that the optimal quantum value of the success probability uniquely certifies that the states prepared by Alice are indeed the set of trine states, and the measurement instrument that Bob uses measures along the directions of the trine spin axes.

Next, we crafted a scenario that allows multiple independent observers to share sequential quantum advantage when they are allowed to implement their respective measurements with varying degrees of unsharpness. In particular, we have explicitly shown that at most two observers can get the quantum advantage if the degree of unsharpness of the first observer's measurement lies inside a specific range of values. Therefore, when both observers get the quantum advantage sequentially, we demonstrated how to certify the range of the unsharpness parameter of the measurement instrument used by the former observer. Most importantly, it is shown that, even when the success probability does not reach the global optimal value, for a given unsharpness parameter, the optimal pair of success probabilities of the respective observers certify the prepared states and measurement instruments.

We further extended our investigation to the case when the optimal value of the success probability for a particular unsharpness parameter is not achieved. In such a scenario, we derive the fidelity bounds of the prepared states and implement measurements to the ideal ones. This enables one to robustly certify the states and measurements that are subjected to imperfections other than unsharp measurements. In the sequential scenario, as each optimal pair of success probability certifies the states and measurement, the parameters that give the bound to the fidelity possess different values for different optimal pairs. We derived the exact forms of parameters as a function of the degree of unsharpness. Thus, the derived trade-off relation between the Fidelity and the success probability provides the robust certification of the states mentioned above and measurements.

To the best of our knowledge, all the works to date considered the robust certification of quantum instruments in a two-party scenario. Nevertheless, as the situation may sometimes demand both the sustainable quantum advantage and certification of the instruments, the robustness analysis in the presence of an unsharp instrument becomes crucial. Our work sheds light in this direction by extending the conventional robustness analysis scheme to the case where the measurements themselves are unsharp along with other experimental imperfections. We hope that future research will find more general methods to handle the unsharpness parameters alongside other experimental imperfections while analyzing the robustness of different certification schemes. Studies along this line would be an interesting avenue for future research.

\section*{Acknowledgments}
Authors acknowledge the anonymous referee for valuable suggestions and comments, which helped to improve the quality of the work.  A.S.S and S. M acknowledge the support from the project DST/ICPS/QuEST/Theme 1/2019/4 and A. K. P acknowledges the support from the project MTR/2021/000908.

\begin{widetext}
\appendix
\section*{Appendix}

\section{Robust self-testing of Alice's preparations}\label{App:A}
We provide the detailed derivation of robustness of preparations of Alice. The average fidelity of the preparations with the ideal states are quantified by
\begin{align}
	\mathcal{S}(\rho_{xa})=\frac{1}{6}\underset{\Lambda}{max}\sum_{x=1}^3 \sum_{a\in\{0,1\}} F(\rho_{xa}^{\ ideal}, \Lambda(\rho_{xa}))
\end{align}
where $\Lambda$ is a CPTP map. The average fidelity is estimated by considering to all the possible channels. The lower bound is derived by minimizing the average fidelity over the set of preparations compatible with  the given value of $\mathcal{A}_B$, i.e,
\begin{align}
	\mathcal{F}(\mathcal{A}_B) = \underset{\rho_{xa}\in R(\mathcal{A}_B)}{min} \mathcal{S}({\rho_{xa}})
\end{align}
where $R(\mathcal{A}_{B})$ is the set of all preparations compatible with the value of $\mathcal{A}_{B}$. As explained in the main text, we write the operator inequality,
\begin{align}
\label{eq:Ineqp}
	K_{xa}(B_1,B_2,B_3)\leq sW_{xa}+t_{xa}(B_1,B_2,B_3)\mathbb{I}
\end{align}
where 
\begin{equation}
\label{eq:Kp}
    K_{xa}(B_1,B_2,B_3) = \Lambda^\dagger(B_1,B_2,B_3)\left[\rho_{xa}^{\ ideal}\right]
\end{equation}
 $\Lambda^\dagger$ is the dual to a quantum channel $ \Lambda$,  and
 \begin{equation}
 \label{eq:Wp}
     W_{xa}=\frac{1}{36}\sum_{y=1}^3 (-1)^{\delta_{x,y} \oplus a}B_y.
 \end{equation}
 Since $W_{xa}$ is a function of $B_1,B_2$ and $B_3$, the quantities $K_{xa}$ and $t_{xa}$ are chosen to be dependent on  $B_1,B_2$ and $B_3$ in order to be in compliance with $W_{xa}$.
 Taking trace over inputs we arrive at
\begin{equation}
	\begin{split}
		\mathcal{S} &\geq \frac{1}{6}\sum_{x=1}^3 \sum_{a\in\{0,1\}} Tr[K_{xa}\rho_{xa}]\geq \frac{s}{6}Tr[\rho_{xa}W_{xa}+t_{xa}\mathbb{I}]=\frac{s}{6}[(\mathcal{A}_B-\half)]+\frac{1}{6}\sum_{x=1}^3 \sum_{a\in\{0,1\}}t_{xa}.
	\end{split}
\end{equation}
Therefore, the lower bound  of average fidelity can be determined by,
\begin{align}
\label{eq:avgfid}
	F(\mathcal{A}_B)=\frac{s}{6}(\mathcal{A}_B-\frac{1}{2}) + t.
\end{align}
where $t= \underset{B_1,B_2,B_3}{min}\frac{1}{6}\sum_{x=1}^3 \sum_{a\in\{0,1\}}t_{xa}(B_1,B_2,B_3)$. Thus, the problem resolves to choose a particular value $s$, and subsequently the values of $t_{xa}$ are minimized with respect to $B_1,B_2$ and $B_3$ for the choice of $s$. \\
We choose a dephasing channel of the form,
\begin{align}\label{eq:channelp}
	\Lambda_\theta(\rho) = \frac{1+c(\theta)}{2}\rho +\frac{1-c(\theta)}{2}\Gamma_\theta \ \rho \ \Gamma_\theta
\end{align}
For    $\theta \in [0,\frac{\pi}{3}]$ ; $\Gamma = \sqt\sigma_z-\half\sigma_x$ and for  $\theta \in (\frac{\pi}{3},\frac{\pi}{2}]$ ; $\Gamma = -\sqt\sigma_z-\half\sigma_x$. The dephasing function $c(\theta) \in [0,1]$ needs to be specified.\\
Therefore, using Eqs.\eqref{eq:Kp} and \eqref{eq:channelp} we find the action of channel in the intervals,\\

for $\theta \in [0,\frac{\pi}{3}]$

\begin{equation}
    \begin{split}
        &K_{10}=\half \qty(\mathbb{I} + \frac{\sqrt{3}(c-1)}{4}\Z + \frac{ (1+3c)}{4}\X),\hspace{10pt}
        K_{11} = \half \qty(\mathbb{I} - \frac{\sqrt{3}(c-1)}{4}\Z - \frac{(1 +3c)}{4}\X), \hspace{10pt}\\ 
        &K_{20}=\half \qty(\mathbb{I} + \frac{\sqrt{3}}{2}\Z -\half\X),\hspace{65pt}
        K_{21}=\half\qty(\mathbb{I} - \frac{\sqrt{3}}{2}\Z +\half\X),\hspace{10pt}\\
        &K_{30}=\half \qty(\mathbb{I} - \frac{\sqrt{3}(1+c)}{4}\Z +\frac{(1 - 3 c)}{4}\X),\hspace{10pt}
        K_{31}=\half \qty(\mathbb{I} + \frac{\sqrt{3}(1+c)}{4}\Z  +\frac{(1 - 3 c)}{4}\X).
    \end{split}
\end{equation}
similarly, for $\theta \in (\frac{\pi}{3},\frac{\pi}{2}]$
\begin{equation}
    \begin{split}
        &K_{10}=\half \qty(\mathbb{I} + \frac{\sqrt{3}(1-c)}{4}\Z + \frac{(1 + 3 c)}{4}\X),\hspace{10pt}K_{11} = \half \qty(\mathbb{I} - \frac{\sqrt{3}(1-c)}{4}\Z - \frac{(1 + 3 c)}{4}\X), \\ &K_{20}=\half \qty(\mathbb{I} + \frac{\sqrt{3}(1+c)}{4}\Z  + \frac{(1 - 3 c)}{4}\X),\hspace{10pt}
        K_{21}=\half \qty(\mathbb{I} - \frac{\sqrt{3}(1+c)}{4}\Z - \frac{(1-3c)}{4}\X),\\ &
        K_{30}=\half \qty(\mathbb{I} - \frac{\sqrt{3}}{4}\Z -\half\X),\hspace{65pt}K_{31}=\half\qty(\mathbb{I} + \frac{\sqrt{3}}{4}\Z +\half\X).
    \end{split}
\end{equation}
Similarly, for finding the expressions for $W_{xa}$ we consider Bob's measurements  are projective and are represented by,
\begin{align}
	B_1 = -\eta_B \sigma_x, \hspace{15pt}B_2 = \eta_B \qty(\cos\theta\sigma_x-\sin\theta\sigma_z), \hspace{15pt}
	B_3 = \eta_B\qty( \cos\theta\sigma_x+\sin\theta\sigma_z).\label{measop}
\end{align}
Hence, by Eq.\eqref{eq:Wp} $W_{xa}$'s can be written as,
\begin{equation}
    \begin{split}\label{eq:wxap}
        W_{10}&=\frac{1}{36}\eta_B \left(\sigma_x+2\cos\theta\sigma_{x}\right),\hspace{10pt}  W_{11}=\frac{1}{36}\eta_B \left(-\sigma_x-2\cos\theta\sigma_{x}\right),\hspace{10pt}
	    W_{20}=\frac{1}{36}\eta_B \left(-\sigma_x+2\sin\theta\sigma_z\right)\\ W_{21}&=\frac{1}{36}\eta_B \left(\sigma_x-2\sin\theta\sigma_z\right),\hspace{10pt}
	    W_{30}=\frac{1}{36}\eta_B \left(-\sigma_x-2\sin\theta\sigma_z\right),\hspace{10pt}W_{31}=\frac{1}{36}\eta_B \left(-\sigma_x-2\sin\theta\sigma_z\right)
    \end{split}
\end{equation}
 Due to the apparent symmetries of $K_{xa}$ and $W_{xa}$, the number of inequalities can be further reduced. If we choose $ t_{10}=t_{11}\equiv t_1$, $t_{20}=t_{21} \equiv  t_2$ and $ t_{30 }=t_{31} \equiv t_3$, in each interval only three inequalities need to be considered.\\
Now, we write the inequalities in the interval $\theta\in[0,\frac{\pi}{3}]$,

\begin{align}
\label{eq:ineqp}
	\half\qty(\mathbb{I}+\frac{\sqrt{3}(c-1)}{4}\Z+\frac{1+3c}{4}\sigma_x)&-\frac{\eta_B s}{36}\left(\sigma_x+2\cos\theta\sigma_{x}\right)-t_1\mathbb{I}\geq 0 \nonumber \\
	\half\qty(\mathbb{I}+\frac{\sqrt{3}}{2}\sigma_z-\frac{1}{2}\sigma_x)&-\frac{\eta_B s}{36}\left(-\sigma_x+2\sin\theta\sigma_z\right)-t_2\mathbb{I}\geq 0,\nonumber \\
	\half\left(\mathbb{I}-\frac{\sqrt{3}(1+c)}{4}\sigma_z-\frac{1(1-3c)}{4}\sigma_x\right)&-\frac{\eta_B s}{36}\left(-\sigma_x-2\sin\theta\sigma_z\right)-t_3\mathbb{I}\geq 0.
\end{align}
and in the interval $\theta\in(\frac{\pi}{3},\frac{\pi}{2}]$
    \begin{align}
        \half\qty(\mathbb{I}-\frac{\sqrt{3}(c-1)}{4}\Z+\frac{1+3c}{4}\sigma_x)&-\frac{\eta_B s}{36}\left(\sigma_x+2\cos\theta\sigma_{x}\right)-t_1\mathbb{I}\geq 0 \nonumber \\
	   \half\qty(\mathbb{I}+\frac{\sqrt{3}(1+c)}{4}\sigma_z-\frac{(1-3c)}{4}\sigma_x)&-\frac{\eta_B s}{36}\left(-\sigma_x+2\sin\theta\sigma_z\right)-t_2\mathbb{I}\geq 0,\nonumber \\
	   \half\left(\mathbb{I}-\frac{\sqrt{3}}{2}\sigma_z-\frac{1}{2}\sigma_x\right)&-\frac{\eta_B s}{36}\left(-\sigma_x-2\sin\theta\sigma_z\right)-t_3\mathbb{I}\geq 0.
    \end{align}

Solving the Eq.\eqref{eq:ineqp} we find the values of $t_1,t_2$ and $t_3$ for the interval $ \theta \in[0,\frac{\pi}{3}] $
        \begin{equation}
            \begin{split}
                t_1 =\text{min}&\Bigg\{\frac{1}{36} \qty(18-\sqrt{81+243c^2-9s\eta_B-27cs\eta_B + 3s^2\eta_B^2-18s\eta_B\cos\theta-54cs\eta_B\cos\theta+4s^2\eta_B^2\cos\theta+ 2s^2\eta_B^2\cos2\theta}),\\ &\quad 
                \frac{1}{36} \qty(18 +\sqrt{81 + 243 c^2 - 9s\eta_B -27cs\eta_B + 3s^2\eta_B^2 -18s\eta_B\cos\theta - 54cs\eta_B\cos\theta + 4s^2\eta_B^2\cos\theta + 2s^2\eta_B^2 \cos2\theta})\bigg\}\\
                t_2 =\text{min}&\Bigg\{\frac{1}{36} \qty(18 -\sqrt{324-18 s\eta_B+3s^2\eta_B^2-2s^2\eta_B^2\cos 2 \theta-36\sqrt{3}s\eta_B\sin\theta}),\\ &\quad
                \frac{1}{36} \qty(18+\sqrt{324-18s\eta_B+3s^2\eta_B^2-2s^2\eta_B^2\cos2\theta-36\sqrt{3}s\eta_B\sin\theta})\Bigg\}\\
                t_3 = \text{min}&\Bigg\{\frac{1}{36} \qty(18-\sqrt{81+243c^2+9s\eta_B-27cs\eta_B+s^2 \eta_B^2-18\sqrt{3}s\eta_B\sin\theta-18\sqrt{3}cs\eta_B\sin\theta+4s^2\eta_B^2\sin^2\theta}),\\ 
                &\quad  
                \frac{1}{36}\qty(18 +\sqrt{81+243c^2+9s\eta_B-27cs\eta_B+s^2\eta_B^2-18\sqrt{3}s\eta_B\sin\theta-18\sqrt{3}cs\eta_B\sin\theta+4s^2\eta_B^2\sin^2\theta})\Bigg\}.
            \end{split}
        \end{equation}
and for the interval $\theta \in (\frac{\pi}{3},\frac{\pi}{2}]$
    \begin{equation}
        \begin{split}
            t_1 = \text{min}&\Bigg\{\frac{1}{36} \qty(18 - \sqrt{81+ 243c^2 -9s\eta_B - 27cs\eta_B + 3s^2\eta_B^2 -18s\eta_B\cos\theta - 54cs\eta_B\cos\theta + 4s^2\eta_B^2\cos\theta + 
            2s^2\eta_B^2\cos2\theta}),\\ 
            &\quad
            \frac{1}{36} \qty(18 + \sqrt{81 + 243c^2 - 9s\eta_B - 27cs\eta_B + 3s^2\eta_B^2 - 18s \eta_B\cos\theta - 54cs\eta_B\cos\theta + 4s^2\eta_B^2\cos\theta + 2s^2\eta_B^2\cos2 \theta})\Bigg\}\\
            t_2 = \text{min}&\Bigg\{\frac{1}{36}\qty(18 - \sqrt{81 + 243c^2 + 9s\eta_B - 27cs\eta_B + 
            s^2\eta_B^2 - 18\sqrt{3}s\eta_B\sin\theta - 18\sqrt{3}cs\eta_B\sin\theta + 4s^2 \eta_B^2\sin^2\theta}),\\
            &\quad
            \frac{1}{36}\qty(18 + \sqrt{81 + 243c^2 + 9s\eta_B - 27cs\eta_B + s^2\eta_B^2 - 18 \sqrt{3}s\eta_B\sin\theta - 18\sqrt{3}cs\eta_B\sin\theta + 4s^2\eta_B^2 \sin^2\theta})\Bigg\}\\
            t_3 = \text{min}&\Bigg\{\frac{1}{36}\qty(18 - \sqrt{324 - 18s\eta_B + 3s^2\eta_B^2 - 2s^2 \eta_B^2\cos2\theta -36\sqrt{3}s \eta_B\sin\theta}),\\
            &\quad
            \frac{1}{36}\qty(18 + \sqrt{324 - 18s\eta_B + 3s^2\eta_B^2 -2s^2\eta_B^2\cos2\theta - 36\sqrt{3}s\eta_B\sin\theta})\Bigg\}
        \end{split}
    \end{equation}          
With the expressions of $K_{xa}$, $W_{xa}$ and $ t$'s, we have constructed the operator inequalities for a given $\theta$ in the form \eqref{eq:Ineqp}. To find lower bound of the average fidelity as given in the Eq.\eqref{eq:avgfid}, we fix the value of s to be $\frac{9}{\eta_B}$.
The dephasing function is assumed to be $c(\theta) = min\{1,\frac{s}{6}\sin\theta\}$ for the interval, $\theta \in [0,\frac{\pi}{3}]$, and $c(\theta) = min\{1,\frac{s}{6}\cos\theta\}$ for the interval $\theta \in (\frac{\pi}{3},\frac{\pi}{2}]$. It is obvious to notice that $c(\theta)\in [0,1]$ and is continuous at $\theta = \frac{\pi}{3}$. By the apparent symmetry three out of six inequalities are considered and substituting the values of $s$ and $c$ we find $t$ to be,
\begin{equation}
   t= \frac{1}{3}\qty(t_1+t_2+t_3) = \half
\end{equation}
This, completely generalizes the robustness analysis, such that the general equation for average fidelity will be
\begin{align}
	\mathcal{F}(\mathcal{A}_B) = \frac{3}{2\eta_B}(\mathcal{A}_B-\frac{1}{2})+\half \label{fidp}.
\end{align}
The Eq.\eqref{fidp} provides the relation between robustness and (non)optimal success probability$ \mathcal{A}_B$. For a fixed value of $\eta_B$, the lower bound $\mathcal{F}(\mathcal{A}_B) = 1$ is attained for the  optimal success probability, $\mathcal{A}_B = \Omega_B$.

\section{Robust self-testing of Measurements}
\label{App:B}
Let us now derive the robustness of measurements for the two sequential observers Bob and Charlie by invoking the similar scheme developed for preparations.To begin with,
the average fidelity of measurements with respect to ideal ones are quantified as,
\begin{align}
	\mathcal{S}'(\{E_{b|y}\}) = \frac{1}{6}\underset{\Lambda}{max}\sum_{y=1}^3\sum_{b\in\{0,1\}}\left(F((E_{b|y})^{\ ideal},\Lambda[E_{b|y}])\right)
\end{align}
where maximization is taken for all possible channels. The lower bound for the average fidelity can be written as,
\begin{align}
	\mathcal{F}'(\mathcal{A}) = \underset{{E_{b|y}}\in R'(\mathcal{A})}{min}\mathcal{S}'(\{E_{b|y}\})
\end{align}
where $\mathcal{F}' $ is minimized over measurements compatible with the value of $\mathcal{A}$. Now, we write the operator inequality in the form
\begin{align}
\label{eq:Ineqm}
	K_{yb}({\rho_{yb}}) \geq sZ_{yb}+t_{yb}\mathbb{I}
\end{align}
where we find that
\begin{equation}
\label{eq:Km}
    K_{yb}=\Lambda^\dagger[(E_{b|y})^{\ ideal}]
\end{equation}
 and 
 \begin{equation}
 \label{eq:Zm}
     Z_{yb}=\frac{1}{18}\sum_{x=1}^3\sum_{a\in\{0,1\}}\rho_{xa}\delta_{l,b}
 \end{equation}
 where $l=\delta_{x,y}\oplus a$.
If we take trace over the input states and minimize $t_{xa}$, we get,
\begin{align}
\label{eq:fidm}
	\mathcal{F}'(\mathcal{A}) = \frac{s}{6}\mathcal{A}+t
\end{align}
where $t=\frac{1}{6}\text{min}_{B_1,B_2,B_3}\sum_{y=1}^3\sum_{b=\{0,1\}}t_{yb}(B_1,B_2,B_3)$. We adopt the same dephasing channel as in Eq.\eqref{eq:channelp} and two intervals of $\theta$
For $\theta \in [0,\frac{\pi}{3}]$, $\theta \in (\frac{\pi}{3},\frac{\pi}{2}]$, we recall that $\Gamma = -\sqt\sigma_z+\half\sigma_x$ and  $\Gamma = \sqt\sigma_z+\half\sigma_x$ respectively.
\subsection{Robust certification of Bob's measurement Bob}\label{rbmb}
As argued in the main text, Bob performs unsharp measurements to extend the quantum advantage to Charlie. It is already derived in the main text that, in oroder to produce optimal success probability Alice's preparation should be pure states. We re-write the states as,
\begin{align}
	\rho_{yb} = \frac{\mathbb{I}+\va{r}_{yb}\vdot\vec{\sigma}}{2}
\end{align}
The explicit form of Bloch vectors are,
\begin{align}
\label{eq:blochvectors}
	\va{r}_{10} &= (1,0,0),&\va{r}_{11} &=(-1,0,0)\nonumber\\
	\va{r}_{20}&=(-\cos\theta,0,\sin\theta),&\va{r}_{21}&=(\cos\theta,0,-\sin\theta)\nonumber\\
	\va{r}_{30}&=(-\cos\theta,0,-\sin\theta),&\va{r}_{31}&=(\cos\theta,0,\sin\theta)
\end{align}

We write the ideal POVMs for Bob as,
\begin{align*}
	\qty(M_{b|1})^{\ ideal}= \frac{\mathbb{I}-(-1)^b\eta_B\sigma_x}{2},\hspace{15pt}
	\qty(M_{b|2})^{\ ideal}&=\frac{1}{2}\left(\mathbb{I}+(-1)^b\eta_B\qty(\frac{1}{2}\sigma_x-\frac{\sqrt{3}}{2}\sigma_z)\right),\hspace{15pt}
	\qty(M_{{b|3}})^{\ ideal}=\frac{1}{2}\left(\mathbb{I}+(-1)^b\eta_B\qty(\frac{1}{2}\sigma_x+\frac{\sqrt{3}}{2}\sigma_z)\right)
\end{align*}
Correspondingly, from Eq.\eqref{eq:Km} Bob's $K_{yb}$ values in the interval $ \theta\in [0,\frac{\pi}{3}]$, can be written as,
\begin{equation}
    \begin{split}
    &K_{10}=\half\qty(\I+\frac{\sqrt{3}(1-c)\eta_B}{4}\Z-\frac{(1+3c)\eta_B}{4}\X),\hspace{10pt} K_{11}=\half\qty(\I-\frac{\sqrt{3}(1-c)\eta_B}{4}\Z+\frac{(1+3c)\eta_B}{4}\X)\\
    &K_{20} = \half\qty(\I-\frac{\sqrt{3}\eta_B}{2}\Z+\half\eta_B\X),\hspace{65pt}K_{21} = \half\qty(\I+\frac{\sqrt{3}\eta_B}{2}\Z-\half\eta_B\X)\\
    &K_{30} = \half\qty(\I+\frac{\sqrt{3}(1+c)\eta_B}{4}\Z+\frac{(3c-1)\eta_B}{4}\X),\hspace{10pt}
    K_{31} = \half\qty(\I-\frac{\sqrt{3}(1+c)\eta_B}{4}\Z-\frac{(3c-1)\eta_B}{4}\X)
    \end{split}
\end{equation}
and for $ \theta\in(\frac{\pi}{3},\frac{\pi}{2}]$,
\begin{equation}
    \begin{split}
    &K_{10}=\half\qty(\I-\frac{\sqrt{3}(1-c)\eta_B}{4}\Z-\frac{(1+3c)\eta_B}{4}\X),\hspace{10pt}K_{11}=\half\qty(\I+\frac{\sqrt{3}(1-c)\eta_B}{4}\Z+\frac{(1+3c)\eta_B}{4}\X)\\
    &K_{20} = \half\qty(\I-\frac{\sqrt{3}(1+c)\eta_B}{4}\Z+\frac{(3c-1)\eta_B}{4}\X),\hspace{10pt}K_{21} = \half\qty(\I+\frac{\sqrt{3}(1+c)\eta_B}{4}\Z-\frac{(3c-1)\eta_B}{4}\X)\\
    &K_{30} = \half\qty(\I+\frac{\sqrt{3}\eta_B}{2}\Z+\half\eta_B\X),\hspace{65pt}
    K_{31} = \half\qty(\I-\frac{\sqrt{3}\eta_B}{2}\Z-\half\eta_B\X)
    \end{split}
\end{equation}
 By using Eq.\eqref{eq:Zm} we derive $Z_{yb}$ in the form,
\begin{equation}
\begin{split}
    &Z_{10} = \qty(\frac{3\I}{2}-\frac{\X+2\cos\theta\X}{2}),\hspace{10pt}
    Z_{11} = \qty(\frac{3\I}{2}+\frac{\X+2\cos\theta\X}{2}),\hspace{10pt}
    Z_{20} = \qty(\frac{3\I}{2}+\frac{\X-2\sin\theta\Z}{2}),\hspace{10pt}\\
    &Z_{21} = \qty(\frac{3\I}{2}+\frac{-\X+2\sin\theta\Z}{2}),\hspace{10pt}
    Z_{30} = \qty(\frac{3\I}{2}+\frac{\X+2\sin\theta\Z}{2}),\hspace{10pt}
    Z_{31} = \qty(\frac{3\I}{2}-\frac{\X+2\sin\theta\Z}{2}).
\end{split}
\end{equation}
From the above expressions for $K_{yb}$ and $Z_{yb}$, we write the  inequalities for the respective intervals, i.e,

for $\theta \in [0,\frac{\pi}{3}]$,
\begin{equation}
\label{eq:ineqm1}
    \begin{split}
        \frac{\I-c\eta_B\X}{2}-&\frac{s}{18}\qty(\frac{3\I}{2}-\frac{\X+2\cos\theta\X}{2})-t_1\I\geq 0\\
        \half\qty(\I-\sqt\eta_B\Z+\half c\eta_B\X)-&\frac{s}{18}\qty(\frac{3\I}{2}+\frac{\X-2\sin\theta\Z}{2})-t_2\I\geq 0\\
        \half\qty(\I+\sqt\eta_B\Z+\half c\eta_B\X)-&\frac{s}{18}\qty(\frac{3\I}{2}+\frac{\X+2\sin\theta\Z}{2})-t_3\I\geq 0
    \end{split}
\end{equation}
for $\theta \in (\frac{\pi}{3},\frac{\pi}{2}]$,
\begin{equation}
\label{eq:ineqm2}
    \begin{split}
        \frac{\I-\eta_B\X}{2}-&\frac{s}{18}\qty(\frac{3\I}{2}-\frac{\X+2\cos\theta\X}{2})-t_1\I\geq 0\\
        \half\qty(\I-\sqt c\eta_B\Z+\half \eta_B\X)-&\frac{s}{18}\qty(\frac{3\I}{2}+\frac{\X-2\sin\theta\Z}{2})-t_2\I\geq 0\\
        \half\qty(\I+\sqt c\eta_B\Z+\half \eta_B\X)-&\frac{s}{18}\qty(\frac{3\I}{2}+\frac{\X+2\sin\theta\Z}{2})-t_3\I\geq 0
    \end{split}
\end{equation}

Solving the inequalities in Eqs. \eqref{eq:ineqm1} and \eqref{eq:ineqm2} we find the values of $t_1, t_2$ and $t_3$ for the interval 
$\theta \in [0,\frac{\pi}{3}]$, 
\begin{equation}
    \begin{split}
        t_1 = \text{min}&\Bigg\{\frac{1}{36}\qty(18-3s-\sqrt{3s^2-9s\eta_B -       27cs\eta_B + 81 \eta_B^2 + 243c^2\eta_B^2 +                   4s^2\cos\theta- 18 s\eta_B \cos\theta - 54cs\eta_B               \cos\theta +2s^2\cos2\theta}),
                \\&\quad
             \frac{1}{36}\qty(18 - 3 s + \sqrt{3s^2 - 9 s \eta_B - 27 c s\eta_B+ 81 \eta_B^2 + 243c^2\eta_B^2 + 4s^2\cos\theta - 18s\eta_B \cos\theta -54cs\eta_B\cos\theta + 2s^2\cos2\theta})\Bigg\}\\
        t_2= \text{min}&\Bigg\{\frac{1}{36}\qty(18 - 3 s - \sqrt{s^2 - 18 s         \eta_B + 324\eta_B^2-36 \sqrt{3}s\eta_B\sin\theta + 4s^2      \sin^2\theta}),\\
             &\quad
            \frac{1}{36}\qty(18 - 3 s + \sqrt{s^2 - 18s\eta_B + 324\eta_B^2 - 36\sqrt{3} s \eta_B \sin\theta + 4 s^2 \sin^2\theta})\Bigg\}\\
        t_3= \text{min}&\Bigg\{\frac{1}{36}\qty(18 - 3s -\sqrt{s^2 + 9s\eta_B - 27cs       \eta_B + 81 \eta_B^2 + 243 c^2 \eta_B^2 - 18 \sqrt{3} s \eta_B       \sin\theta - 18\sqrt{3} c s \eta_B \sin\theta + 4 s^2 \sin^2\theta}),
        \\&\quad
        \frac{1}{36}\qty(18 - 3 s + \sqrt{s^2 + 9 s \eta_B - 27 c s \eta_B + 81 \eta_B^2 + 243 c^2 \eta_B^2 - 18 \sqrt{3} s \eta_B \sin\theta - 
      18 \sqrt{3} c s \eta_B \sin\theta + 4 s^2 \sin^2\theta})\Bigg\}
    \end{split}
\end{equation}
 and for the interval $\theta \in (\frac{\pi}{3},\frac{\pi}{2}]$,
\begin{equation}
    \begin{split}
        t_1 = \text{min}&\Bigg\{\frac{1}{36}\qty(18 - 3s -\sqrt{3s^2 - 9s\eta_B - 27 cs\eta_B + 81\eta_B^2 + 243c^2\eta_B^2 + 4s^2\cos\theta - 18s\eta_B \cos\theta - 54 c s \eta_B \cos\theta + 2s^2\cos2\theta}),\\
        &\quad
        \frac{1}{36}\qty(18 - 3 s + \sqrt{3s^2 - 9s\eta_B - 27cs\eta_B + 81 \eta_B^2 + 243c^2\eta_B^2 + 4s^2\cos\theta - 18s\eta_B\cos\theta - 54cs \eta_B\cos\theta + 2s^2\cos 2 \theta})\Bigg\}\\
        t_2 = \text{min}&\Bigg\{\frac{1}{36}\qty(18 - 3 s - \sqrt{s^2 + 9 s \eta_B - 27cs\eta_B + 81\eta_B^2 + 243c^2\eta_B^2 -18\sqrt{3}s\eta_B\sin\theta - 18\sqrt{3}cs\eta_B\sin\theta +4s^2 \sin^2\theta}),\\  
        &\quad
        \frac{1}{36}\qty(18 - 3 s + \sqrt{s^2 + 9 s \eta_B - 27 c s \eta_B + 81 \eta_B^2 + 243 c^2 \eta_B^2 - 18 \sqrt{3} s \eta_B \sin\theta - 
        18 \sqrt{3} c s \eta_B \sin\theta + 4 s^2 \sin^2\theta})\Bigg\}\\
        t_3 = \text{min}&\Bigg\{\frac{1}{36}\qty(18 - 3 s - \sqrt{s^2 - 18 s \eta_B + 324 \eta_B^2 -36 \sqrt{3} s \eta_B \sin\theta + 
        4 s^2 \sin^2\theta}),\\
        &\quad
        \frac{1}{36}\qty(18 - 3 s + \sqrt{s^2 - 18 s \eta_B + 324 \eta_B^2 - 
        36\sqrt{3} s \eta_B \sin\theta + 4 s^2 \sin^2\theta})\Bigg\}
    \end{split}
\end{equation}
Now, for a given value of $\theta$, with expressions for $t$, $K_{yb}$, and $Z_{yb}$  we have constructed the operator inequalities of the form \eqref{eq:Ineqm} . To get the lower bound on $\mathcal{F}'$ as given in Eq.\eqref{eq:fidm} we choose s=9 and minimizing $t$ with respect to $B_1$, $B_2$ and $B_3$, by the apparent symmetry, $ t = \frac{t_1+t_2+t_3}{3}$. Simple calculation gives $t- \qua - \frac{\eta_B}{2}$. Thus, we write the average fidelity of measurements for Bob as,
\begin{equation}
    \mathcal{F}'(\mathcal{A_B}) = \frac{3}{2}\mathcal{A}_B +\qua -\frac{\eta_B}{2}
\end{equation}
which is same as Eq.\eqref{eq:fidmb} in the main text.The robustness of Bob's measurement is generalized for the (non)optimal success probability and the choice $\eta_B$, quantified in terms of average fidelity. The lower bound of the average fidelity is achieved when $\mathcal{A}_B=\Omega_B$.
\subsection{Robust certification of measurements for Charlie}
In the same way, we devised the calculations for robustness for Bob by the same approach; here we derive the same for Charlie as well. We find the values of $K_{yb}$ for Charlie in the interval

 $ \theta\in[0,\frac{\pi}{3}]$, the values of $K_{yb}$'s can be written as,
\begin{equation}
    \begin{split}
    &K_{10}=\half\qty(\I+\frac{\sqrt{3}(1-c)}{4}\Z-\frac{(1+3c)}{4}\X),\hspace{10pt}
    K_{20} = \half\qty(\I-\frac{\sqrt{3}}{2}\Z+\half\X),\hspace{10pt}
    K_{30} = \half\qty(\I+\frac{\sqrt{3}(1+c)}{4}\Z+\frac{(3c-1)}{4}\X)\\
    &K_{11}=\half\qty(\I-\frac{\sqrt{3}(1-c)}{4}\Z+\frac{(1+3c)}{4}\X),\hspace{10pt}
    K_{21} = \half\qty(\I+\frac{\sqrt{3}}{2}\Z-\half\X),\hspace{10pt}
    K_{31} = \half\qty(\I-\frac{\sqrt{3}(1+c)}{4}\Z-\frac{(3c-1)}{4}\X)
    \end{split}
\end{equation}
Similarly, For $ \theta\in(\frac{\pi}{3},\frac{\pi}{2}]$, the values of $K_{yb}$'s can be written as,
\begin{equation}
    \begin{split}
    &K_{10}=\half\qty(\I-\frac{\sqrt{3}(1-c)}{4}\Z-\frac{(1+3c)}{4}\X),\hspace{10pt}
    K_{20} = \half\qty(\I-\frac{\sqrt{3}(1+c)}{4}\Z+\frac{(3c-1)}{4}\X),\hspace{10pt}
    K_{30} = \half\qty(\I+\frac{\sqrt{3}}{2}\Z+\half\X)\\
    &K_{11}=\half\qty(\I+\frac{\sqrt{3}(1-c)}{4}\Z+\frac{(1+3c)}{4}\X),\hspace{10pt}
    K_{21} = \half\qty(\I+\frac{\sqrt{3}(1+c)}{4}\Z-\frac{(3c-1)}{4}\X),\hspace{10pt}
    K_{31} = \half\qty(\I-\frac{\sqrt{3}}{2}\Z-\half\X)
    \end{split}
\end{equation}
To find the expression of $Z_{yb}$, the states provided by Bob are simplified to the form,
\begin{align}
	\rho_{yb} = \frac{\mathbb{I}+\gamma\va{r}_{yb}\vdot\vec{\sigma}}{2}
\end{align}
where, $\gamma = \frac{1+\sqrt{1-\eta_B^2}}{2}$ and $\va{r}_{yb}$ is same as defined in Eq.\eqref{eq:blochvectors}. We find the expressions for $Z_{yb}$ in terms of $\rho_{yb}$ as,
\begin{align}
	&Z_{10} = \frac{1}{18}\qty(\frac{3\I}{2}-\frac{\gamma}{2}\X-\gamma\cos\theta \X ), \hspace*{10pt}	Z_{11} = \frac{1}{18}\qty(\frac{3\I}{2}+\frac{\gamma}{2}\X+\gamma\cos\theta \X )\nonumber,\hspace*{10pt} 
	Z_{20}= \frac{1}{18}\qty(\frac{3\I}{2}+\frac{\gamma}{2}\X-\gamma\sin\theta \Z ),\\\hspace*{20pt} &Z_{21}= \frac{1}{18}\qty(\frac{3\I}{2}-\frac{\gamma}{2}\X+\gamma\sin\theta \Z ),\hspace*{10pt}
	Z_{30}= \frac{1}{18}\qty(\frac{3\I}{2}+\frac{\gamma}{2}\X+\gamma\sin\theta \Z ),\hspace*{20pt} Z_{31}= \frac{1}{18}\qty(\frac{3\I}{2}-\frac{\gamma}{2}\X-\gamma\sin\theta \Z )\label{stat}
\end{align}

The operator inequalities for 
For the interval $\theta\in[0,\frac{\pi}{3}]$,
\begin{align}
	&\half\qty(\I+\frac{\sqrt{3}(1-c)}{4}\Z-\frac{(1+3c)}{4}\X)-\frac{s}{18}\qty(\frac{3\I}{2}-\frac{\gamma}{2}\X-\gamma\cos\theta \X ) -t_1\I \geq0 \nonumber\\
	&\half\qty(\I-\frac{\sqrt{3}}{2}\Z+\half\X)-\frac{s}{18}\qty(\frac{3\I}{2}+\frac{\gamma}{2}\X-\gamma\sin\theta \Z )-t_2\I\geq0\nonumber\\
	&\half\qty(\I+\frac{\sqrt{3}(1+c)}{4}\Z+\frac{(3c-1)}{4}\X)-\frac{s}{18}\qty(\frac{3\I}{2}+\frac{\gamma}{2}\X+\gamma\sin\theta \Z )-t_3\I\geq0.
\end{align}
For the interval $\theta\in(\frac{\pi}{3},\frac{\pi}{2}]$,
\begin{align}
	&\half\qty(\I-\frac{\sqrt{3}(1-c)}{4}\Z-\frac{(1+3c)}{4}\X)-\frac{s}{18}\qty(\frac{3\I}{2}-\frac{\gamma}{2}\X-\gamma\cos\theta \X ) -t_1\I \geq0 \nonumber\\
	&\half\qty(\I-\frac{\sqrt{3}(1+c)}{4}\Z+\frac{(3c-1)}{4}\X)-\frac{s}{18}\qty(\frac{3\I}{2}+\frac{\gamma}{2}\X-\gamma\sin\theta \Z )-t_2\I\geq0\nonumber\\
	&\half\qty(\I+\frac{\sqrt{3}}{2}\Z+\half\X)-\frac{s}{18}\qty(\frac{3\I}{2}+\frac{\gamma}{2}\X+\gamma\sin\theta \Z )-t_3\I\geq0.
\end{align}
Solving the inequalities the results for $ t$'s in the interval 

Solving the inequalities the results for $\theta\in[0,\frac{\pi}{3}]$will be,
\begin{equation}
    \begin{split}
        t_1 = \text{min}&\Bigg\{\frac{1}{36} (18 - 3 s - \sqrt{
    81 + 243 c^2 - 9 s \gamma - 27 c s \gamma + 
     3 s^2 \gamma^2 - 18 s \gamma \cos\theta - 
     54 c s \gamma \cos\theta + 4 s^2 \gamma^2 \cos\theta + 
     2 s^2 \gamma^2 \cos2 \theta}), \\ &\quad
  \frac{1}{36} (18 - 3 s + \sqrt{
     81 + 243 c^2 - 9 s \gamma - 27 c s \gamma + 
      3 s^2 \gamma^2 - 18 s \gamma \cos\theta - 
      54 c s \gamma \cos\theta + 
      4 s^2 \gamma^2 \cos\theta + 
      2 s^2 \gamma^2 \cos2 \theta})\Bigg\}\\
        t_2 = \text{min}&\Bigg\{\frac{1}{36} \qty(18 - 3 s - \sqrt{324 - 18 s \gamma + s^2 \gamma^2 - 36 \sqrt{3} s \gamma \sin\theta + 4 s^2 \gamma^2 \sin^2\theta}), 
        \\ & \qquad
        \frac{1}{36} \qty(18 - 3 s + \sqrt{324 - 18 s \gamma + s^2 \gamma^2 - 
        36 \sqrt{3} s \gamma \sin\theta + 4s^2 \gamma^2 \sin^2\theta})\Bigg\}\\
        t_3 = \text{min}&\Bigg\{\frac{1}{36} \qty(18 - 3 s - \sqrt{81 + 243 c^2 + 9 s \gamma - 27 c s \gamma + s^2 \gamma^2 - 18 \sqrt{3} s \gamma \sin\theta - 
        18 \sqrt{3} c s \gamma \sin\theta + 4s^2 \gamma^2 \sin^2\theta}), \\ &\qquad
        \frac{1}{36} (18 - 3 s + \sqrt{81 + 243 c^2 + 9 s \gamma - 27 c s \gamma + s^2 \gamma^2 - 18 \sqrt{3} s \gamma \sin\theta - 18 \sqrt{3} c s \gamma \sin\theta + 4s^2 \gamma^2 \sin^2\theta})\Bigg\}
    \end{split}
\end{equation}
Solving the inequalities the results for $\theta\in(\frac{\pi}{3}, \frac{\pi}{2}]$will be,
\begin{equation}
    \begin{split}
        t_1 = \text{min}&\Bigg\{\frac{1}{36} \qty(18 - 3     s - \sqrt{81 + 243 c^2 - 9 s \gamma - 27 c s \gamma + 3 s^2 \gamma^2 - 18 s \gamma \cos \theta - 54 c s \gamma \cos \theta + 4 s^2 \gamma^2 \cos \theta + 2 s^2 \gamma^2 \cos 2 \theta}),\\&\quad
        \frac{1}{36} \qty(18 - 3 s + \sqrt{81 + 243 c^2 - 9 s \gamma - 27 c s \gamma + 3 s^2 \gamma^2 - 18 s \gamma \cos \theta - 54 c s \gamma \cos \theta + 4 s^2 \gamma^2 \cos \theta + 2 s^2 \gamma^2 \cos 2 \theta})\Bigg\}\\
        t_2 = \text{min}&\Bigg\{\frac{1}{36} \qty(18 - 3 s - \sqrt{81 + 243 c^2 + 9 s \gamma - 27 c s \gamma + s^2 \gamma^2 - 18 \sqrt{3} s \gamma \sin\theta - 18 \sqrt{3} c s \gamma \sin\theta + 4 s^2 \gamma^2 \sin^2\theta}),\\ &\quad 
        \frac{1}{36} \qty(18 - 3 s + \sqrt{81 + 243 c^2 + 9 s \gamma - 27 c s \gamma + s^2 \gamma^2 - 18 \sqrt{3} s \gamma \sin\theta - 18 \sqrt{3} c s \gamma \sin\theta + 4 s^2 \gamma^2 \sin^2\theta})\Bigg\}\\
        t_3 = \text{min}&\Bigg\{\frac{1}{36} \qty(18 - 3 s - \sqrt{324 - 18 s \gamma + s^2 \gamma^2 - 36 \sqrt{3} s \gamma \sin\theta + 4 s^2 \gamma^2 \sin^2\theta}),\\ &\quad 
        \frac{1}{36} \qty(18 - 3 s + \sqrt{324 - 18 s \gamma + s^2 \gamma^2 - 36 \sqrt{3} s \gamma \sin\theta + 4 s^2 \gamma^2 \sin^2\theta})\Bigg\}
    \end{split}
\end{equation}

 Now we have constructed the operator inequality for a given value of $\theta$. To get the lower bound on average fidelity of measurement $(\mathcal{F}')$ as given in Eq.\eqref{eq:fidm}, we fix $s=\frac{54}{7\gamma-1}$ and the subsequent minimization gives $t=\frac{(9 - 6 \gamma)}{(2 - 14 \gamma)}$ which provides a complete generalized analysis of robustness based on $\mathcal{A}_C$.
 Therefore, the average fidelity of measurements for Charlie can be written as,

\begin{align}
	 \label{fidm}
		\mathcal{F}'(\mathcal{A}_C) &= \frac{9}{7\gamma-1}\mathcal{A}_C + \frac{(9 - 6 \gamma)}{(2 - 14 \gamma)}.
\end{align}
 The equation above is the same as Eq. \eqref{eq:cff} in the main text and describes the variation of average fidelity with success probability  $\mathcal{A}_C$ of Charlie for a particular choice of $\eta_{B}$. Lower bound of average fidelity, $\mathcal{F}'(\mathcal{A}) = 1$ is attained only when $\mathcal{A}_C = \Omega_C$.
\end{widetext}


\begin{thebibliography}{99}
	\bibitem{bell} J.S. Bell, On the Einstein Podolsky Rosen paradox, \href{https://doi.org/10.1103/PhysicsPhysiqueFizika.1.195}{Physics, {\bf 1}, 195 (1964)}.

	
	\bibitem{brunnerrev} N. Brunner, D. Cavalcanti, S. Pironio, V. Scarani and S. Wehner, Bell nonlocality, \href{https://doi.org/10.1103/RevModPhys.86.419}{Rev. Mod. Phys. 86, 419 (2014)}.
	
		\bibitem{bar05} J. Barrett, L. Hardy and A. Kent, No Signaling and Quantum Key Distribution, \href{https://journals.aps.org/prl/abstract/10.1103/PhysRevLett.95.010503}{Phys. Rev. Lett. 95, 010503(2005)}.
			\bibitem{acin06} A. Acin, N. Gisin and L. Masanes, From Bell’s Theorem to
			Secure Quantum Key Distribution, \href{https://journals.aps.org/prl/abstract/10.1103/PhysRevLett.97.120405}{Phys. Rev. Lett. 97, 120405 (2006)}.
			\bibitem{acin07} A. Acin, N. Brunner, N. Gisin, S. Massar, S. Pironio and and V.
			Scarani, Device-Independent Security of Quantum Cryptography against Collective Attacks, \href{https://journals.aps.org/prl/abstract/10.1103/PhysRevLett.98.230501}{ Phys. Rev. Lett. 98, 230501 (2007)}. 
			\bibitem{pir09} S. Pironio, A. Acin, N. Brunner, N. Gisin, S. Massar and V.
			Scarani, Device-independent quantum key distribution secure against collective attacks, \href{https://iopscience.iop.org/article/10.1088/1367-2630/11/4/045021/meta}{New J. Phys. 11, 045021 (2009)}. 
			
			\bibitem{col06} R. Colbeck, Quantum and relativistic protocols for secure multi-party computation, Ph.D. thesis, University of Cambridge (2006);  \href{https://arxiv.org/abs/0911.3814}{arXiv:0911.3814v2}. 
			\bibitem{pir10} S. Pironio, et al., Random numbers certified by Bell’s theorem, \href{https://www.nature.com/articles/nature09008}{Nature (London) 464, 1021 (2010)}.
			\bibitem{nieto} O. Nieto-Silleras, S. Pironio and J. Silman, Using complete measurement statistics for optimal device-independent randomness evaluation, \href{https://iopscience.iop.org/article/10.1088/1367-2630/16/1/013035}{New J. Phys. 16, 013035 (2014)}. 
			\bibitem{col12}R. Colbeck and R. Renner, Free randomness can be amplified, \href{https://www.nature.com/articles/nphys2300}{Nat. Phys. 8, 450 (2012)}.
			\bibitem{wehner} S. Wehner, M. Christandl and A.C. Doherty, Lower bound on the dimension of a quantum system given measured data, \href{https://journals.aps.org/pra/abstract/10.1103/PhysRevA.78.062112}{Phys. Rev. A 78, 062112 (2008)}. 
			\bibitem{gallego} R. Gallego, N. Brunner, C. Hadley and A. Acin, Device-Independent Tests of Classical and Quantum Dimensions,  \href{https://journals.aps.org/prl/abstract/10.1103/PhysRevLett.105.230501}{Phys. Rev. Lett. 105, 230501 (2010)}. 
			\bibitem{ahrens} J. Ahrens, P. Badziag, A. Cabello and M. Bourennane, Experimental device-independent tests of classical and quantum dimensions,  \href{https://www.nature.com/articles/nphys2333}{ Nat. Phys. 8, 592 (2012)}. 
			\bibitem{brunnerprl13} N. Brunner, M. Navascues and T. Vertesi, Dimension Witnesses and Quantum State Discrimination,  \href{https://journals.aps.org/prl/abstract/10.1103/PhysRevLett.110.150501}{Phys. Rev. Lett. 110,  150501 (2013)}.
			
			\bibitem{bowler} J. Bowles, M. T. Quintino and N. Brunner, Certifying the Dimension of Classical and Quantum Systems in a Prepare-and-Measure Scenario with Independent devices, \href{https://journals.aps.org/prl/abstract/10.1103/PhysRevLett.112.140407}{Phys. Rev. Lett. 112, 140407 (2014)}.
			\bibitem{sik16prl}J. Sikora, A. Varvitsiotis and Z. Wei, Minimum Dimension of a Hilbert Space Needed to Generate a Quantum Correlation, \href{https://journals.aps.org/prl/abstract/10.1103/PhysRevLett.117.060401}{Phys. Rev. Lett., 117, 060401 (2016)}.
			
			\bibitem{cong17} W. Cong, Y. Cai, J-D. Bancal and V. Scarani, Witnessing Irreducible Dimension, \href{https://journals.aps.org/prl/abstract/10.1103/PhysRevLett.119.080401}{Phys. Rev. Lett. 119, 080401 (2017)}. 
			
			\bibitem{pan2020} A. K. Pan and S. S. Mahato, device-independent certification of the Hilbert-space dimension using a family of Bell expressions, \href{https://doi.org/10.1103/PhysRevA.102.052221}{Phys. Rev. A 102, 052221 (2020)}.
			
			\bibitem{complx1} H. Buhrman, R. Cleve, S. Massar, and R. de Wolf, Nonlocality and communication complexity, \href{https://journals.aps.org/rmp/pdf/10.1103/RevModPhys.82.665}{Phys. Rev. Lett. 114, 250401 (2015)}.
	
	        \bibitem{buhrman16}{H. Buhrman etal., Quantum communication complexity advantage implies violation of a Bell inequality} \href{https://www.pnas.org/doi/10.1073/pnas.1507647113}{ 10.1073/pnas.1507647113}.
	

	
	\bibitem{KSpeker68} S. Kochen and E. P. Specker, The Problem of Hidden Variables in Quantum Mechanics, \href{https://www.jstor.org/stable/24902153?seq=1#metadata_info_tab_contents} { J. Math. Mech. 17, 59 (1967)}.
	
	
	\bibitem{bell66} J.S Bell, On the problem of hidden variables in quantum mechanics, \href{https://journals.aps.org/rmp/abstract/10.1103/RevModPhys.38.447} {Rev. Mod. Phys. 38, 447 (1966)}.
	
	\bibitem{spekkens05} R. W. Spekkens, Contextuality for preparations, transformations, and unsharp measurements, \href{https://journals.aps.org/pra/abstract/10.1103/PhysRevA.71.052108} {Phys. Rev. A 71, 052108 (2005)}.
	
	\bibitem{harrigen} N. Harrigan and R. W. Spekkens , Incompleteness, and the Epistemic View of Quantum States, \href{https://link.springer.com/article/10.1007} {Foundations of Physics, 40 (2010)}.
	
	\bibitem{ballentine} L. E. Ballentine, Ontological Models in Quantum Mechanics: What do they tell us?, \href{https://arxiv.org/pdf/1402.5689v1.pdf}{	arXiv:1402.5689 (2014)}.
	
	\bibitem{banik14} M. Banik, S. S. Bhattacharya, S. K. Choudhary, A. Mukherjee, and A. Roy , Ontological Models, Preparation Contextuality and Nonlocality, \href{https://link.springer.com/article/10.1007/s10701-014-9839-4}{Foundations of Physics volume 44, pages1230–1244 (2014)}.
	
	
	
	
	
	
	
	\bibitem{spek09} R. W. Spekkens,  D. H. Buzacott, A. J.  Keehn, B. Toner and G. J. Pryde,  Preparation contextuality powers parity-oblivious multiplexing, \href{https://journals.aps.org/prl/abstract/10.1103/PhysRevLett.102.010401}{Phys. Rev. Lett. 102, 010401 (2009)} .
    
    \bibitem{hameedi} A. Hameedi, A. Tavakoli, B. Marques, and M. Bourennane, Communication Games Reveal Preparation Contextuality, \href{https://journals.aps.org/prl/abstract/10.1103/PhysRevLett.119.220402}{Phys. Rev. Lett. 119, 220402 (2017)}.
	
	\bibitem{ghorai} S. Ghorai and A. K. Pan, Optimal quantum preparation contextuality in an $n$-bit parity-oblivious multiplexing task, \href{https://journals.aps.org/pra/abstract/10.1103/PhysRevA.98.032110}{Phys. Rev. A 98, 032110 (2018)}.
	
	\bibitem{saha19b} D. Saha and A. Chaturvedi, Preparation contextuality as an essential feature underlying quantum communication advantage, \href{https://journals.aps.org/pra/abstract/10.1103/PhysRevA.100.022108}{Phys. Rev. A 100, 022108 (2019)}.
	
	\bibitem{pan19} A. K. Pan, Revealing universal quantum contextuality through communication games, \href{https://www.nature.com/articles/s41598-019-53701-5}{Sci Rep 9, 17631 (2019)}.

    \bibitem{kumari2019}  A. Kumari and A.K. Pan, Sharing nonlocality and nontrivial preparation contextuality using the same family of Bell expressions, \href{https://journals.aps.org/pra/abstract/10.1103/PhysRevA.100.062130}{ Phys. Rev. A 100, 062130 (2019)}.
    
     \bibitem{schmid18}  D. Schmid and R. W. Spekkens, Contextual Advantage for State Discrimination, \href{https://journals.aps.org/prx/abstract/10.1103/PhysRevX.8.011015}{Phys. Rev. X 8, 011015 (2018)}.
     
      \bibitem{mukherjee22} S. Mukherjee, S. Naonit and A.K. Pan, Discriminating three mirror symmetric states with restricted contextual advantage, \href{https://journals.aps.org/pra/abstract/10.1103/PhysRevA.106.012216?ft=1}{Phys. Rev. A 106, 012216 (2022)}.
     
     
      \bibitem{flatt22} K. Flatt, H. Lee, C. R. i Carceller, J. B. Brask, and J. Bae, Contextual Advantages and Certification for Maximum-Confidence Discrimination, \href{https://journals.aps.org/prxquantum/cited-by/10.1103/PRXQuantum.3.030337}{PRX Quantum 3, 030337 (2022)}.
     
     \bibitem{lostaglio20} M. Lostaglio, and G. Senno, Contextual advantage for state-dependent cloning, \href{https://quantum-journal.org/papers/q-2020-04-27-258/}{Quantum 4, 258 (2020)}.
     

     
     
     
    \bibitem{supicrev}  I. \v{S}upi{\'{c}} and J. Bowles, Self-testing of quantum systems: a review, \href{https://doi.org/10.22331/q-2020-09-30-337}{Quantum 4, 337 (2020)}.
    
    \bibitem{kaniewski}J. Kaniewski, Analytic and Nearly Optimal Self-Testing Bounds for the Clauser-Horne-Shimony-Holt and Mermin Inequalities, \href{https://link.aps.org/doi/10.1103/PhysRevLett.117.070402}{Phys. Rev. Lett. 117, 070402 (2016)}.

    \bibitem{chsh} J. F. Clauser, M. A. Horne, A. Shimony, and R. A. Holt, Proposed Experiment to Test Local Hidden-Varxable Theories, \href{https://journals.aps.org/prl/references/10.1103/PhysRevLett.23.880} {Phys. Rev. Lett. 24, 549 (1970)}.
   

	
	
	\bibitem{tava2018}A. Tavakoli, J. Kaniewski, T. V\'{e}rtesi, D. Rosset and N. Brunner, Self-testing quantum states and measurements in the prepare-and-measure scenario. \href{https://link.aps.org/doi/10.1103/PhysRevA.98.062307}{Phys. Rev. A 98, 062307 (2018)}.
	
		
	\bibitem{tava21} H. Anwer, N. Wilson, R. Silva, S. Muhammad, A. Tavakoli, and M. Bourennane, Noise-robust preparation contextuality shared between any number of observers via unsharp measurements.\href{https://quantum-journal.org/papers/q-2021-09-28-551/}{Quantum 5, 551 (2021)}.
	
	\bibitem{farkas2019} M. Farkas and J. Kaniewski, Self-testing mutually unbiased bases in the prepare-and-measure scenario, \href{https://journals.aps.org/pra/abstract/10.1103/PhysRevA.99.032316}{Phys. Rev. A 99, 032316 (2019)}.

	\bibitem{pawlowski11} M. Paw\l{}owski and N. Brunner, Semi-device-independent security of one-way quantum key distribution,  \href{https://journals.aps.org/pra/abstract/10.1103/PhysRevA.84.010302}{Phys. Rev. A 84, 010302(R) (2011)}.
	
	\bibitem{miklin2020}   N. Miklin, J. J. Borka\l{}a and M. Paw\l{}owski, Semi-device-independent self-testing of unsharp measurements,  \href{https://journals.aps.org/prresearch/abstract/10.1103/PhysRevResearch.2.033014}{ Phys. Rev. Research 2, 033014 (2020)}.
	\bibitem{anwar2020} H. Anwer, S. Muhammad, W. Cherifi, N. Miklin, A. Tavakoli and M. Bourennane, Experimental Characterization of Unsharp Qubit Observables and Sequential Measurement Incompatibility via Quantum RAC, \href{https://journals.aps.org/prl/abstract/10.1103/PhysRevLett.125.080403}{Phys. Rev. Lett. 125, 080403 (2020)}. 
	\bibitem{fole2020} G. Foletto, L. Calderaro, G. Vallone and P. Villoresi, Experimental demonstration of sequential quantum random access
	codes, \href{https://journals.aps.org/prresearch/abstract/10.1103/PhysRevResearch.2.033205}{Phys. Rev. Research 2, 033205 (2020)}.
	\bibitem{tava20exp}  A. Tavakoli, M. Smania T. V\'{e}rtesi, N. Brunner and M. Bourennane, Self-testing nonprojective quantum measurements in prepare-and-measure experiments, \href{https://advances.sciencemag.org/content/6/16/eaaw6664.abstract}{Science Advances 6, 16 (2020)}.
	\bibitem{sumit21} S. Mukherjee and A. K. Pan, Semi-device-independent certification of multiple unsharpness parameters through sequential measurements, \href{https://journals.aps.org/pra/abstract/10.1103/PhysRevA.104.062214}{Phys. Rev. A 104, 062214 (2021)}.
	
	\bibitem{buhrman10} H. Buhrman, R. Cleve, S. Massar, R. de Wolf, Nonlocality and communication complexity. \href{https://journals.aps.org/rmp/abstract/10.1103/RevModPhys.82.665}{Rev Mod. Phys.,  82, 665 (2010)}.
	

	\bibitem{tava2020} A. Tavakoli, E. Z. Cruzeiro, J. Bohr Brask, N. Gisin and N. Brunner, Informationally restricted quantum correlations, \href{https://quantum-journal.org/papers/q-2020-09-24-332/}{ Quantum, 4, 332 (2020)}.
	
	\bibitem{lsw} Y.C. Liang, R. W.Spekkens, H. M. Wiseman, Specker’s parable of the overprotective seer: A road to contextuality, nonlocality and complementarity, \href{https://www.sciencedirect.com/science/article/pii/S0370157311001517?}{Physics Reports, 506, Issues 1–2 (2011)}.
		
	\bibitem{xiao} Ya Xiao, Xin-Hong Han, Xuan Fan, Hui-Chao Qu, and Yong-Jian Gu, Widening the sharpness modulation region of an entanglement-assisted sequential quantum random access code: Theory, experiment, and application, \href{https://doi.org/10.1103/PhysRevResearch.3.023081} {Phys. Rev. Research 3 023081 (2021)}.
	
	\bibitem{silva2015} R. Silva, N. Gisin, Y. Guryanova and S. Popescu,Multiple Observers Can Share the Nonlocality of Half of an Entangled Pair by Using Optimal Weak Measurements, \href{https://journals.aps.org/prl/abstract/10.1103/PhysRevLett.114.250401}{Phys. Rev. Lett. 114, 250401 (2015)}.
	\bibitem{sasmal2018} S. Sasmal , D. Das , S. Mal and A. S. Majumdar, Steering a single system sequentially by multiple observers, \href{https://journals.aps.org/pra/abstract/10.1103/PhysRevA.98.012305}{Phys. Rev. A 98, 012305  (2018)}.
	
	\bibitem{bera2018}  A. Bera, S. Mal, A. Sen(De) and U. Sen, Witnessing bipartite entanglement sequentially by multiple observers, \href{https://journals.aps.org/pra/abstract/10.1103/PhysRevA.98.062304}{ Phys. Rev. A 98, 062304 (2018)}.
	\bibitem{brown2020}  P. J. Brown and R. Colbeck, Arbitrarily Many Independent Observers Can Share the Nonlocality of a Single Maximally Entangled Qubit Pair, \href{https://journals.aps.org/prl/abstract/10.1103/PhysRevLett.125.090401}{Phys. Rev. Lett. 125, 090401  (2020)}.
	
	\bibitem{Zhang21} T. Zhang and S-M. Fei, Sharing quantum nonlocality and genuine nonlocality with independent observables, \href{https://journals.aps.org/pra/abstract/10.1103/PhysRevA.103.032216}{Phys. Rev. A 103, 032216  (2021)}.
	
	\bibitem{mohan2019}   K. Mohan, A. Tavakoli and N. Brunner, Sequential random access codes and self-testing of quantum measurement instruments, \href{https://iopscience.iop.org/article/10.1088/1367-2630/ab3773}{ New J. Phys. 21 083034 (2019)}.
	
	

	

	\bibitem{vonneumann}  J. von Neumann, \textit{Mathematical Foundations of Quantum Mechanics}, \href{https://press.princeton.edu/books/hardcover/9780691178561/mathematical-foundations-of-quantum-mechanics}{Princeton Univ. Press (1955)}.
	
	\bibitem{busch}  P. Busch, Unsharp reality and joint measurements for spin observables,  \href{https://journals.aps.org/prd/abstract/10.1103/PhysRevD.33.2253}{Phys. Rev. D. 33, 2253 (1986)}.
	
		\bibitem{ovrcomplete01} Roger B. M. Clarke, V. M. Kendon, A. Chefles, S. M. Barnett, E. Riis, and M. Sasaki, Experimental realization of optimal detection strategies for overcomplete states, \href{https://journals.aps.org/pra/abstract/10.1103/PhysRevA.64.012303}{Phys. Rev. A 64, 012303 (2001)}.
	
	\bibitem{mon1} T. J. Osborne, and F. Verstraete, General Monogamy Inequality for Bipartite Qubit Entanglement, \href{https://journals.aps.org/prl/abstract/10.1103/PhysRevLett.96.220503}{Phys. Rev. Lett. 96, 220503  (2006)}.
	\bibitem{std1} J. Bergou, E. Feldman, and M. Hillery, Extracting Information from a Qubit by Multiple Observers: Toward a Theory of Sequential State Discrimination,  \href{https://journals.aps.org/prl/abstract/10.1103/PhysRevLett.111.100501}{Phys. Rev. Lett. 111, 100501  (2013)}.
	\bibitem{std2} D. Fields, R. Han, M. Hillery and J. A. Bergou, Extracting unambiguous information from a single qubit by sequential observers,  \href{https://journals.aps.org/pra/abstract/10.1103/PhysRevA.101.012118}{Phys. Rev. A 101, 012118 (2020)}.
	
	
	
	\bibitem{pan21} A. K. Pan, Oblivious communication game, self-testing of projective and nonprojective measurements, and certification of randomness,  \href{https://journals.aps.org/pra/abstract/10.1103/PhysRevA.104.022212}{Phys. Rev. A, 104, 022212 (2021)}.
	

	
	
	
	
	
	
	
	
	
	
	
	
\end{thebibliography}
\end{document}